\documentclass[twocolumn]{article}

\usepackage{algorithm}
\usepackage{algorithmic}
\usepackage{amsmath}
\usepackage{amsthm}
\usepackage{amssymb}
\usepackage{graphicx}
\usepackage{epstopdf}
\usepackage{balance}
\usepackage{times}
\usepackage{xspace}
\usepackage{url}
\usepackage{xcolor}
\usepackage{multirow}
\usepackage{fullpage}

\newcommand{\comment}[1]{}
\newcommand{\sem}{INSTRUCT}
\newcommand{\figwidth}{0.52\columnwidth}

\title{
\sem{}: Space-Efficient Structure for Indexing and Complete Query
Management of String Databases
}

\author{
\begin{tabular}{cc}
	Sourav Dutta & Arnab Bhattacharya \\
	\url{sodutta3@in.ibm.com} & \url{arnabb@iitk.ac.in} \\
	{IBM Research India,} & {Dept. of Computer Science and Engineering,} \\
	{New Delhi} & {Indian Institute of Technology, Kanpur,} \\
	{India} & {Kanpur, India} \\
\end{tabular}
}

\date{}

\begin{document}

\maketitle 

\begin{abstract}
	The tremendous expanse of search engines, dictionary and thesaurus storage,
	and other text mining applications, combined with the popularity of readily
	available scanning devices and optical character recognition tools, has
	necessitated efficient storage, retrieval and management of massive text
	databases for various modern applications.  For such applications, we
	propose a novel data structure, \sem{}, for efficient storage and management
	of sequence databases.  Our structure uses bit vectors for reusing the
	storage space for common triplets, and hence, has a very low memory
	requirement.  \sem{} efficiently handles prefix and suffix search queries in
	addition to the exact string search	operation by iteratively checking the
	presence of triplets.  We also propose an extension of the structure to
	handle substring search efficiently, albeit with an increase in the space
	requirements.  This extension is important in the context of trie-based
	solutions which are unable to handle such queries efficiently.  We perform
	several experiments portraying that \sem{} outperforms the existing
	structures by nearly a factor of two in terms of space requirements, while
	the query times are better.  The ability to handle insertion and deletion of
	strings in addition to supporting all kinds of queries including exact
	search, prefix/suffix search and substring search makes \sem{} a complete
	data structure. \\
\end{abstract}


\section{Introduction}
\label{sec:intro}

Efficient manipulation of large sets of strings has emerged as a basic
requirement for a growing number of applications including search
engines~\cite{web-27}, port cataloging on the web~\cite{ip-17}, dictionary and
thesaurus support~\cite{trie-13,trie-28}, news archive, document repository,
mining XML databases~\cite{xml-10,xml-9}, searching reserved words in a
compiler~\cite{compl-15}, automaton
searching~\cite{text-30}, text compression~\cite{txt-16}, and indexing huge
databases.  To enhance the performance of retrieval and update queries,
mechanisms reducing the storage space requirement, making them in-memory if
possible, are critical.  With the tremendous improvement in scanning and optical
character recognition technologies along with the efforts in
internationalization and localization, the amount of textual data is beginning
to explode.  Storing such a vast amount of data itself poses a big problem.  The
further requirement of in-memory index structures for fast
look-ups~\cite{comp-25} calls for a compressed representation of even the index
structure.  

Tries~\cite{trie-1} and similar constructs try to achieve this by storing each
character as a node in a tree and reusing some of the prefix nodes.  Since each
string is represented as a path from the root to a leaf, the memory requirement
is large~\cite{trie-19,tree-20}, thereby limiting their application for large
text databases.  Compact tries~\cite{tree-20} and suffix
trees~\cite{suf-31,trie-21} aim to alleviate this problem by reusing the storage
space of the common prefix or suffix of the strings.  However, once two strings
differ in a single character, their paths differ, and they are stored separately
even though the rest may be the same.  In other words, these structures do not
aim to reuse the characters forming the strings.  As all strings are composed of
a defined set of characters, reusing the storage space for common characters
promises to provide the most compressed form of representation.  This redundancy
linked with the need for extreme space-efficient index structures motivated us
to develop \emph{\sem{} (INdexing STrings by Re-Using Common Triplets)}.

With the size of databases breaking the barrier of terabytes, efficient data
mining operations call for fast techniques for tackling prefix, suffix and
substring searches.  Prefix and suffix search queries allow context-based data
retrieval.  Data compression techniques, as in the sorting stage of
Burrows-Wheeler transform~\cite{bw-34} also utilize such searches.  Even data
clustering algorithms, like suffix tree clustering used in search engines make
use of efficient suffix searching.  Pattern or substring search is an important
query operation in large genome and text data storage, and is used in software
maintenance~\cite{softmain-34} and text editing among other related fields.  

We show that \sem{} efficiently handles such search queries, thereby making it a
complete indexing structure.  While the experiments show that \sem{} does not
achieve industrial-scale (orders of magnitude) speed-ups over the competing
structures, we feel that the ability of \sem{} to handle all string operations
at a better or equal cost makes it a comprehensive structure for string
databases. 

In a nutshell, our contributions are as follows:
\begin{enumerate}
	\item We have designed an intelligent structure \sem{} for sequence indexing
		that reuses the storage space for common characters.
	\item We have depicted how different operations such as insertion and
		searching, including prefix, suffix and substring searching, can be
		efficiently supported by our structure.
	\item We have shown that \sem{} outperforms the existing structures by up to
		a factor of two in memory requirements while maintaining better or
		comparable running times for searching and insertion.
\end{enumerate}

The paper is organized as follows.  Section~\ref{sec:survey} provides a glimpse
of the existing data structures for string management.
Section~\ref{sec:structure} defines the structure of \sem{}.  Algorithms for
insertion, searching, etc. using \sem{} are presented and analyzed in
Section~\ref{sec:algo}.  Section~\ref{sec:expts} reports the experimental
results before Section~\ref{sec:concl} concludes.

\section{Related Work}
\label{sec:survey}

Although hashing~\cite{hash-29,hash-26} provides the fastest way of indexing
keys, the fact that the size of the hash table depend heavily on the data
collision rate, coupled with no reuse of common character storage, often compels
disk accesses, thereby limiting its efficiency.  Moreover, it does not support
efficient prefix, suffix or substring search operations.
Tries~\cite{trie-13,trie-1} are tree-like structures that reuse the storage
space for common prefixes, by storing each subsequent character separately as a
node.  Compact tries~\cite{trie-32,tree-20} fold the tree path leading up to a
single leaf node, i.e., a single suffix, into a single node.  The suffix
tree~\cite{suf-31,trie-21,suffix_tree-2} and prefix tree~\cite{trie-1,prog-14}
respectively collapse the common suffix or prefix into single nodes, but with
the increase in the number of unique keys stored, the length of such common
suffixes and prefixes decreases, whereby the structures degenerate.  Patricia
tries~\cite{pat-24} extend the concept of folding used by compact tries to
single-branch nodes even within the tree structure to increase space efficiency,
but uses optimizations to restrict false positive query results.  Ternary search
trees (TST)~\cite{tst-5,trie-6} are 3-way tree structure with each branching
node replaced by a binary search tree.  This optimization makes the TSTs require
less space than the standard tries~\cite{trie-22}, but also make them much
slower.  VLC-tries~\cite{vlc-3} and LZ-tries~\cite{lz-4} do reduce the storage
space required, but have significantly complex structures and procedures for
querying, which are difficult to implement.  VLC-trie uses the
divide-and-conquer method to obtain a partition of the edges of the trie into
levels that are compressed.  Dictionary compression methods like RLE,
front-compression, and the LZ family~\cite{compr-33} represent data in
compressed form, and  use Patricia tries, prefix trees, and LZ tries
respectively.  However, these methods have highly involved insertion procedure,
and dynamic operations are not well supported.  For example, the basic trie
structure does not support efficient substring searching, while prefix and
suffix trees are biased towards only a subset of the family of search
procedures.  Several other similar structures such as the suffix array cater to
this end.  However, \sem{} inherently allows efficient search procedures for all
the above methods with lower memory requirements.  Burst trie~\cite{burst-7}
stores keys in buckets indexed by trie-like paths and dynamically splits (or
bursts) the buckets during insertion.  Although it is currently the most
space-efficient structure~\cite{hat-8}, its performance varies widely with the
heuristic for the choice of parameters governing the bursting of the overflowing
nodes.  B-tries~\cite{burst-8} provide a disk version of burst tries.

The common space inefficiency of all these structures arise from the lack of
reuse of storage for the individual characters forming the keys.  \sem{}
utilizes just a single node for each triplet of characters, and maps each
triplet of a key into the corresponding node.  It, thus, forms an efficient
in-memory data structure.  The keys are stored based on the 3-grams~\cite{ngram}
present, with a unit window shift to obtain the next trigram.  Indexing with
\sem{} is therefore closely related to that using n-gram indexing~\cite{n-gram}.
In \sem{}, a set bit represents all strings containing the triplet, and there is
no need to merge the results as in the case of n-gram indexing.  This makes
\sem{} simpler and faster.  Further, the optimizations achieved by reusing the
space, and bit vectors that allow efficient pruning along with the robust range
of operations supported makes \sem{} more attractive than the simple n-gram
indexing.

\section{Structure of \sem{}}
\label{sec:structure}

We assume that keys (or equivalently, strings or words) that need to be indexed
are sequences of characters from a alphabet of size $k$.  We also assume that
the maximum length of any key is at most $l$.  For example, in an English
dictionary, $k = 26$ and $l = 29$\footnote{The longest non-technical word in
English is \emph{floccinaucinihilipilification}
(\url{http://en.wikipedia.org/wiki/Longest_word_in_English}).}  If there any $m$
keys, the total number of characters in the database is $d \leq ml$.

The \sem{} structure comprises a collection of $k$ \emph{nodes}, each
corresponding to a particular character of the alphabet.  Each node in turn
comprises a $k \times k$ \emph{matrix}.  A \emph{cell} in the matrix corresponds
to a particular sequence of 3 characters.  We refer to this 3-character set as a
\emph{triplet} or a \emph{3-gram}.  The cell in the node $c_1$ at row $c_2$ and
at column $c_3$ represents the triplet $c_1c_2c_3$ where $c_i$ denotes a
character from the alphabet.  When a particular triplet is present in a key in
the database, the corresponding cell is marked.

However, a triplet may occur at different offsets in a key.  It is thus
beneficial to include this position information in the index.  To enable
indexing of positions, a cell is further broken up into an array of $l$
elements, corresponding to $l$ positions where a triplet can occur in a
key\footnote{Only $l-2$ positions are needed, as there can be a maximum of $l-2$
triplets from a key of length $l$.  However, we ignore this to simplify the
discussion.}.  When a triplet occurs, only the corresponding element is marked.
This, we call the \emph{position} array.

Although \sem{} can naturally adapt to dynamically increasing string lengths,
fixing the length initially makes the representation simple as then all the
structures---nodes, matrices, arrays---become regular arrays of fixed size, and
the \sem{} structure can be very efficiently implemented as a 4-dimensional
\emph{bit array} where the bits can be directly accessed and the bit operations
easily performed.  

\begin{figure}[t]
	\begin{center}
		\includegraphics[width=0.60\columnwidth]{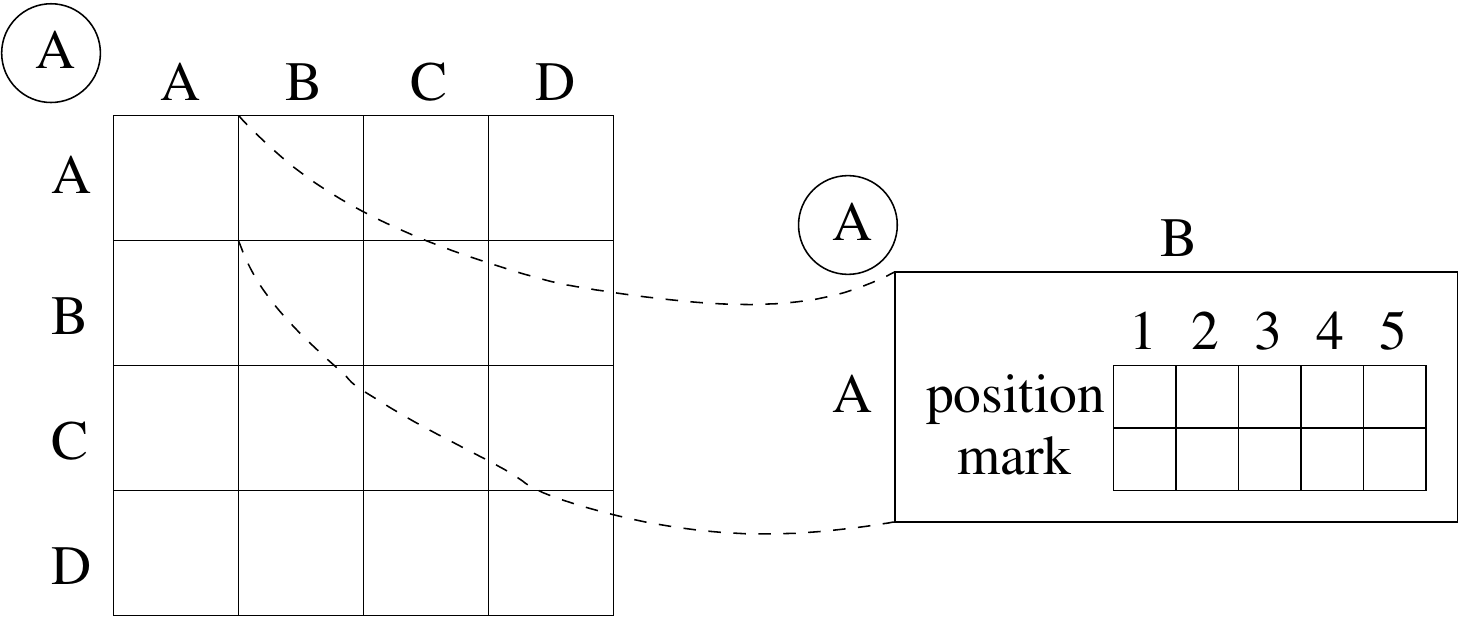}
		\caption{Internal structure of a matrix and a cell.}
		\label{fig:cell}
	\end{center}
\end{figure}

When a particular bit, at say, node $c_1$, row $c_2$, column $c_3$, and position
$w$ is set, it indicates that there exists a key in the database with the
triplet $c_1c_2c_3$ at position $w$.  Figure~\ref{fig:cell} shows the details of
a matrix and a cell where $k=4$ and $l=5$.  The \sem{} structure can be viewed
as a hash table of triplets with position information.

However, unfortunately, the \sem{} structure itself is not enough to
disambiguate between all the keys in a database.  To explain this, consider the
following situation.  Suppose only the keys `ABCA' and `DBCD' are present in a
database.  A search on the key `ABCD' will now be successful as all triplets of
`ABCD', i.e., both `ABC' and `BCD' are marked in \sem{}, and at correct
positions, too!  The problem is that since only triplets are indexed, the
history regarding the original string to which the triplet was a part of, gets
lost.

To alleviate the problem, \sem{} utilizes another $l$-element bit array called
\emph{mark} in each cell, similar to the \emph{position} array.  A bit in the
mark array gets set for a triplet \emph{only} when it is the last triplet in a
key.  Figure~\ref{fig:cell} shows how the mark array is maintained inside a
cell.  When a mark bit is set, a container is allocated that stores \emph{all}
keys that end with the triplet at the position corresponding to the mark
element.  The container may be a lexicographically ordered list or a tree-based structure.  We discuss the
choice of container later.  For the above example, the container for `BCD' will
only include the key `DBCD', and therefore, a search for `ABCD' will fail.  The
containers may also be stored in the disk, if necessary, and pointers to them
are maintained within \sem{}.  For searching and insertion, only the required
container needs to be brought into memory.

For non-string databases, \sem{} can be used to index the primary keys, while
the pointers will be to the buckets containing the complete data stored on disk.

The total space requirement of \sem{} is, thus, only $2k^3l$ bits in addition to
the actual keys (and associated objects).  For the English dictionary, this
translates to only 125\,kB.  It is interesting to observe that for a given value
of $k$ and $l$, all possible permutations of characters up to length $l$ (i.e.,
$k + k^2 + \dots + k^l = O(k^{l+1})$) can be represented in \sem{} with the same
memory requirement.  This feature is quite novel, and makes \sem{} extremely
robust and space-efficient as compared to other structures.  Further, bit
implementation allows simple bit operations such as AND, RIGHT SHIFT, etc. in
the algorithms for searching and insertion (presented in
Section~\ref{sec:algo}), thereby making them extremely efficient.

For extreme pathological cases, where the database is so huge that even this
index cannot be accommodated in the main memory, the individual nodes of \sem{}
can be easily stored in the disk, as they are independently processed for the
different triplets.  The nodes (and corresponding containers) can be dynamically
loaded.  Using various caching and paging policies, the performance in such
situations can be quite efficient.  We do not assume such cases in this paper.

\section{Algorithms}
\label{sec:algo}

\subsection{Insertion}
\label{sec:insertion}

The insertion procedure into \sem{} is based on repeatedly setting the correct
position bits based on all the triplets present in the key.  For the triplet
$c_1c_2c_3$ at position $w$ in the key, the bit in the position array indexed by
node $c_1$, row $c_2$, column $c_3$ and position $w$ is set.  If this is the
last triplet of the key, i.e., $c_3$ is the last character, then the
corresponding mark bit is also set.  If there is a container already pointed to
by the bit (as there may be other keys in the database ending with $c_1c_2c_3$
at $w$), the new key is inserted into the container.  If there is no such
container, a new one is allocated and the key is inserted.  The setting of the
bits can be efficiently implemented using bit-wise operators with appropriate
bit masks.  Without loss of generality, we consider that unique keys are
inserted into \sem{} as primary keys are never duplicated.  In the situation
where keys may be duplicated, the containers will be implemented as a tree-based
structure, and the insertion procedure will be replaced by a search-and-insert
procedure where a key is searched initially, and is inserted only if it is
absent.  

For keys of size $1$ and $2$, we maintain a special container, the size of which
is bounded by $k + k^2$.  This handles the boundary conditions where no proper
triplet can be formed.

\begin{figure}[t]
	\begin{center}
		\begin{tabular}{ccc}
		\includegraphics[width=0.38\columnwidth]{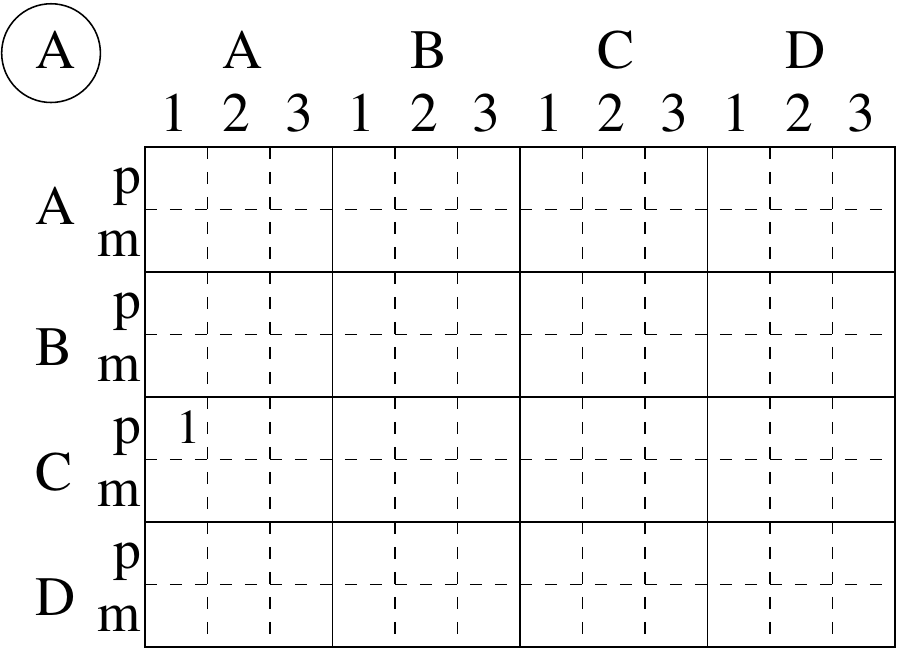} & & 
		\includegraphics[width=0.50\columnwidth]{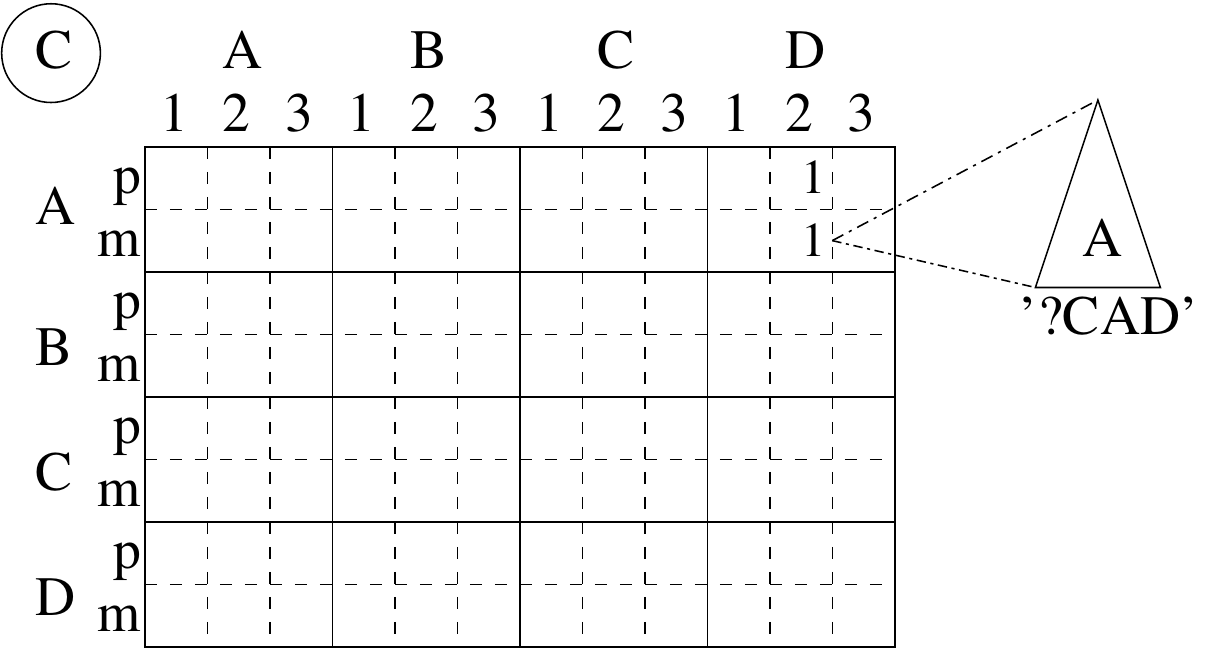} \\
		(a) & & (b)
		\end{tabular}
		\caption{Insertion of key `ACAD': (a) first triplet (`\underline{ACA}D') and (b) last triplet (`A\underline{CAD}').}
		\label{fig:insert}
	\end{center}
\end{figure}

Consider inserting the key `ACAD'.  The first triplet is `ACA'.  Following the
algorithm, $position[A][C][A][1]$ is set (Figure~\ref{fig:insert}(a)).  In the
next step, both $position[C][A][D][2]$ and $mark[C][A][D][2]$ are set
(Figure~\ref{fig:insert}(b)).  Since the key has ended, a container is
allocated.  All keys of the form `?CAD' are indexed in this container, where `?'
stands for any character.  \emph{As a further space optimization, since the last
triplet, i.e., `CAD' is common for all keys in the container, only the rest,
i.e., `A', is stored.}  

Inserting a key of length $n$ requires setting $n-2$ bits corresponding to the
triplets in the key.  Since array addressing takes constant time, the time taken
in this phase is $O(n)$.  After the mark bit is set, the key is inserted into
the container.  Thus, the total time to insert a key is $O(n)$ + (time to insert
in container).  The latter time depends on the nature of the container as well
as its size.  If the container is a list, e.g., a linked list or a dynamic
array, insertion can be achieved in $O(1)$ time.  If, on the other hand, the
container is organized as a tree-structure, e.g., a balanced binary search tree
(BST), insertion takes $O(\log s)$ time where $s$ is the size of the container.

\subsection{Searching}
\label{sec:searching}

Searching a key in \sem{} follows the same procedure as insertion.  For every
triplet in the key, the corresponding bit at the particular position is checked
(again we use masks and bit-wise operators for this purpose).  For the final
triplet, the mark bit is also checked.  If any such bit is not set, then the key
cannot be in the database, and the search is terminated.  So, there are no false
negatives.

However, even if all such bits are set, the container pointed to by the mark bit
needs to be searched, as the bits may be set due to the presence of the key
(successful search) or may be due to the presence of other keys in the database
that together happen to contain all the triplets at the right positions
(unsuccessful search).  Thus, a subsequent search in the container is required
to resolve between the two cases.  In Section~\ref{sec:anal_searching}, we
estimate the probability of such a false positive.

Consequently, in the worst case, the time for searching a key of length $n$ is
$O(n)$ + (time to search in container).  If the container is a linked list of
size $s$, the latter time is $O(s)$; if it is a BST, the time is $O(\log s)$.

\begin{figure}[t]
	\begin{center}
		\includegraphics[width=0.99\columnwidth]{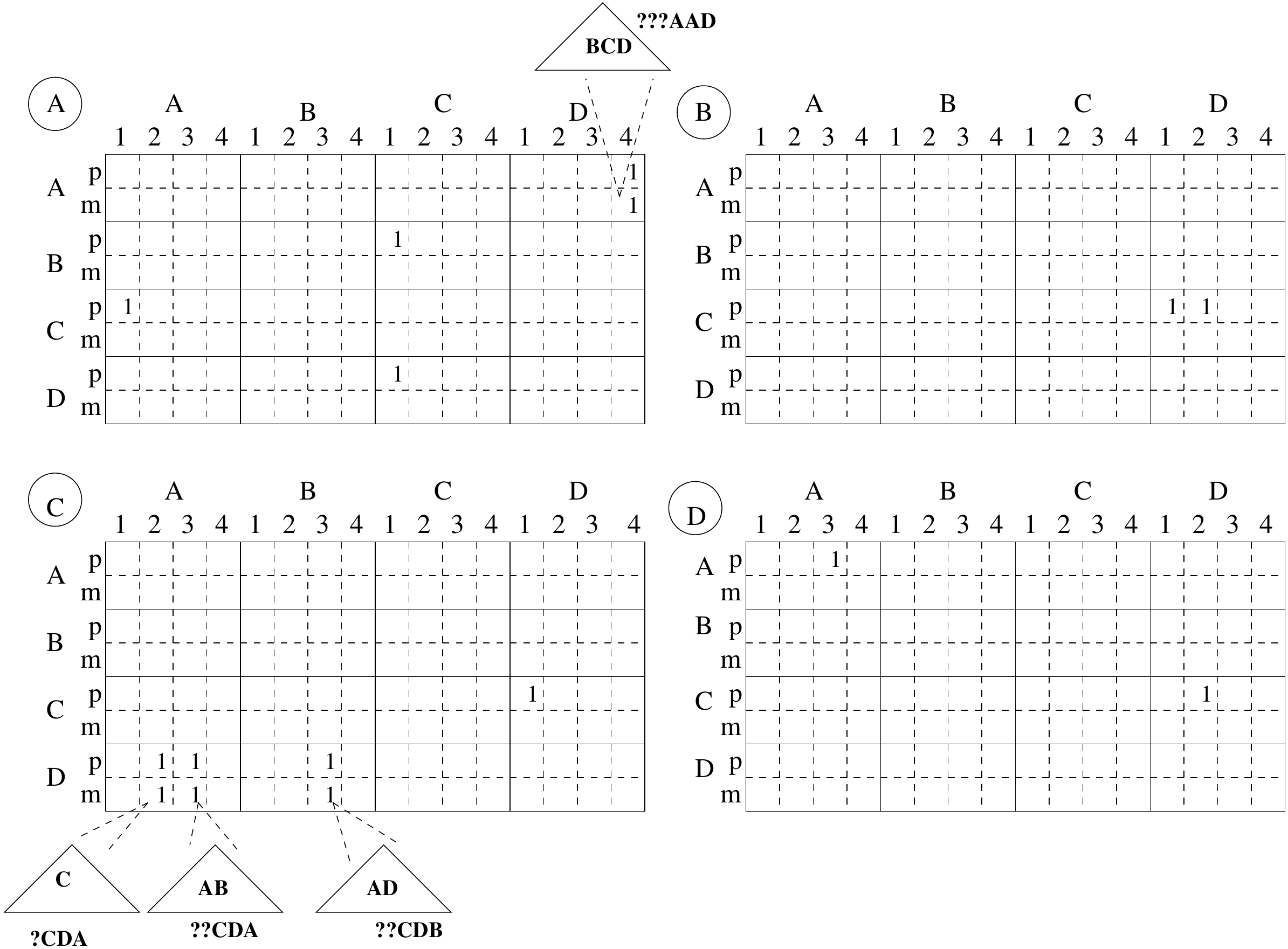}
		\caption{Example of \sem{} storing the keys `ABCDA', `ADCDB', `CCDA',
		and `BCDAAD'.}
		\label{fig:search}
	\end{center}
\end{figure}

Figure~\ref{fig:search} shows the snapshot of a \sem{} structure storing the
keys `ABCDA', `ADCDB', `CCDA', and `BCDAAD'.  Assume that the key `ADCDB' is
queried.  For the first triplet `ADC', we obtain the position bit vector from
the corresponding node.  It must contain a set bit at the first position.  Since
that is the case here, the position vector for the next triplet `DCD' is
checked, which has the second bit set.  Moving forward, for the last and final
triplet `CDB', both the position and the mark vectors contain a set bit at the
third position. Thus, the container corresponding to the mark bit is searched.
It should be noted that the last triplet is not stored in the containers as it
can be obtained from the position of the mark bit.  Consequently, the string
`ADCDB' is reported as present.

If the key `DCDB' is queried, the searching stops at the first step since the
position vector for `DCD' does not contain a set bit at position $1$.  A more
interesting case is for the key `ADCDA'.  All the triplets have the position
bits correctly set and the container corresponding to the final triplet `CDA' is
searched.  This is an example of a false positive search using the \sem{} index
only, as finally the container search returns a negative answer.

The searching algorithm can also follow another strategy.  Only the mark bit
corresponding to the last triplet is examined.  If it is not set, the search
fails.  Otherwise, the container is directly searched without checking the bits
for the other triplets.  This avoids traversing the length of the search key
(i.e., the $O(n)$ time in the total cost).  However, the chance that an
unsuccessful search is terminated early is eliminated.  On the other hand, for a
successful search, this is always a better strategy.  We call this the
\emph{direct search} strategy as opposed to the \emph{index search} strategy
otherwise.

\subsection{Analysis of searching}
\label{sec:anal_searching}

We now analyze the chance of an unsuccessful search being terminated early, and
use that to devise the optimum search strategy.  An unsuccessful search key of
length $n$ will be searched in a container if and only if for every triplet and
position the key generates, the corresponding position bits are set, i.e., for
every triplet $c_1c_2c_3$ at position $w$, there is another key in the database
with the same triplet $c_1c_2c_3$ at the same position $w$.

Since not all keys may be of length $w$, we denote the number of keys in the
database having a length of \emph{at least} $w$ by $f(w)$ and the probability
that at least $1$ out of $m$ keys in the database contains character $c_1$ at
position $w$ by $P_w$.  Assuming all the characters to be equi-probable, i.e.,
the probability of occurrence of a character at any particular position is
$1/k$, we get,
\begin{align}
	\label{eq:single}
	P_{w} &= 1 - P(\text{no key contains } c_1) \nonumber \\
	&= 1 - (P(\mbox{key contains character other than } c_1) )^{f(w)} \nonumber \\
	&= 1 - \left(1 - 1/k\right)^{f(w)}
\end{align}
The probability that a triplet appears at the position $w$ is then the product
of the three individual probabilities (since the corresponding events are
independent):
\begin{align}
	\label{eq:triplet}
	P_{w,3} &= P_{w}.P_{w+1}.P_{w+2} \nonumber \\
	&= \left(1 - \left(1 - 1/k\right)^{f(w)}\right).\left(1 - \left(1 - 1/k\right)^{f(w+1)}\right). \nonumber \\
	&\qquad \qquad \qquad \qquad \qquad \left(1 - \left(1 - 1/k\right)^{f(w+2)}\right) \nonumber \\
	&\simeq 1 - \sum_{i=w}^{w+2}\left(1 - 1/k\right)^{f(i)} \text{[ignoring higher order terms]}
\end{align}

Eq.~\eqref{eq:triplet} provides a way to compute the probability of all $n-2$
triplets appearing at positions $1, \dots, n-2$.  The last triplet, however,
must also be the last triplet in some other key of the same length.  Denoting
the number of database keys that has a length of \emph{exactly} $w$ by $g(w)$,
Eq.~\eqref{eq:single} can be modified as:
\begin{align}
	\label{eq:single_end}
	P_{w_e} &= 1 - \left(1 - 1/k\right)^{g(w)}
\end{align}

Consequently, Eq.~\eqref{eq:triplet} can be modified to:
\begin{align}
	\label{eq:triplet_end}
	P_{w_e,3} &\simeq 1 - \sum_{i=w}^{w+1}\left(1 - 1/k\right)^{f(i)} -
	\left(1 - 1/k\right)^{g(w+2)}
\end{align}

The occurrence of two consecutive triplets is \emph{not} independent as they
share two characters.  However, for simplifying the calculations, we assume that
the events are independent.\comment{In such scenario, we over-estimate the
existence of a false search in the containers.  In reality, the pruning will be
far better.  We establish this by experimental results shown in
Section~\ref{sec:expts}.}  With this assumption, the probability $P_n$ that all
the triplets of the search key of length $n$ are present in the database can be
estimated as
\begin{align}
	\label{eq:entire}
	P_{n} &= \left(\prod_{j=1}^{n-3} P_{j,3} \right). P_{n-2_e,3} \nonumber \\
	&= \prod_{j=1}^{n-3} \left(1 - \sum_{i=j}^{j+2}\left(1 -
	1/k\right)^{f(i)}\right) . \nonumber \\
	&\qquad \qquad \left(1 - \sum_{i=n-2}^{n-1}\left(1 -
	1/k\right)^{f(i)} - \left(1 - 1/k\right)^{g(n)}\right)
	\nonumber \\
	&\simeq 1-\sum_{j=1}^{n-2} \sum_{i=j}^{j+2}\left(1-
	1/k\right)^{f(i)}+\left(1-1/k\right)^{f(n)} \nonumber \\
	&\quad - \left(1 - 1/k\right)^{g(n)} \text{[ignoring higher order terms]}
\end{align}

Since each of the $f(i)$ and $g(i)$ terms are bounded by $m$, $P_n$ can be upper
bounded as follows:
\begin{align}
	\label{eq:entire_bound}
	P_{n} &\leq 1-\sum_{j=1}^{n-2} \sum_{i=j}^{j+2}\left(1-
	1/k\right)^{m} = 1-3(n-2)\left(1-1/k\right)^{m}
\end{align}

Eq.~\eqref{eq:entire_bound} can be used to determine the optimal search
strategy.  Assume that searching for a key through \sem{} takes $T_s$ time and
that through a container takes $T_c$ time.  For an unsuccessful key of length
$n$, the search is terminated using the \sem{} index structure with probability
$(1 - P_{n})$.  Otherwise, with probability $P_{n}$, the container is searched
as well.  Thus, the expected searching time for this \emph{index search}
strategy is
\begin{align}
	\label{eq:index}
	T_{i} = (1 - P_n) T_s + P_n (T_s + T_c)
\end{align}

The alternate \emph{direct search} strategy first checks whether the mark bit is
set for the last (i.e., $(n-2)^{\text{th}}$) triplet, and only if so, searches
the associated container.  The expected time, thus, is
\begin{align}
	\label{eq:index2}
	T_{d} = P_{n-2_e,3} T_c
\end{align}

Thus, it is beneficial to search through \sem{} when
\begin{align}
	\label{eq:index_wins}
	T_i &\leq T_d \nonumber \\
	\text{or, } T_s &\leq (P_{n-2_e,3} - P_n) T_c
\end{align}

Using Eq.~\eqref{eq:entire_bound} and replacing $f(i)$, $g(i)$, etc. in
Eq.~\eqref{eq:triplet_end} by $m$,
\begin{align}
	\label{eq:wins}
	T_s/T_c \leq 3(n-3)\left(1 - 1/k\right)^m
\end{align}

When the length of an unsuccessful search key, $n$, increases, the probability
of the search being pruned by \sem{} increases, as it is less likely that all
the triplets will be present at precisely the right positions.  On the other
hand, when the number of keys, $m$, is very large, due to the large number of
triplets, it becomes more likely that there exists a triplet in the database at
a particular position.  As a result, searching through \sem{} wastes time as
there will be little pruning.  The size of the alphabet, $k$, has an opposing
effect.  When the number of possible characters increase, it is less likely that
a triplet will be repeated in the database, thereby making the chance of pruning
an unsuccessful search higher.  Eq.~\eqref{eq:wins} confirms these behaviors.
Section~\ref{sec:expts} experimentally establishes them.

\subsection{Suffix Searching}
\label{sec:suffix}

The suffix search procedure is almost the same as the exact key search, except
for one crucial difference.  For an exact string search, since the length of the
search key is known, only the particular position bit is checked in the mark
array corresponding to the last triplet of the key.  A suffix, on the other
hand, can end at any length and one particular mark bit cannot be checked.  If,
however, the lengths are known, then the suffix can be easily searched by
iterating over all such possible lengths.  The trick, therefore, is figuring out
these lengths efficiently.

Suppose the query suffix is $c_1c_2 \dots c_f$.  For the last triplet, i.e.,
$c_{f-2}c_{f-1}c_f$, we check at what positions it ends in the mark array.  If
there is a mark bit set at position $p$, it means that there exists a key in the
database that ends at position $p$ with the triplet $c_{f-2}c_{f-1}c_f$.  We
next check the previous triplet $c_{f-3}c_{f-2}c_{f-1}$ in the position array.
If a key contains both the triplets, then the position of the last triplet must
be \emph{exactly} one more than the position of the last but one triplet.
Thus, for every set bit at position $p$ in the mark array, if there is no set
bit at position $p-1$ in the previous array, there cannot be a key ending at
position $p$ containing both the triplets $c_{f-2}c_{f-1}c_f$ and
$c_{f-3}c_{f-2}c_{f-1}$.  Hence, the query cannot be a suffix ending at position
$p$, and the position $p$ can be removed from the list of possible positions.
We continue in this fashion for all the triplets in the suffix.  For all the
mark bits that survive this pruning, we do a search in the corresponding
containers.

For efficiency purposes, the above operations are performed using bit vectors.
The mark and position arrays are all bit vectors.  To obtain all the $p-1$
positions from the mark vector, it is RIGHT SHIFT-ed by one bit.  The resulting
vector is then AND-ed with the position vector of the previous triplet to obtain
the new list of positions.  The RIGHT SHIFT and AND operations are done at most
$f-2$ times for a suffix of length $f$.

Consider searching the suffix `BCDA' in the \sem{} structure shown in
Figure~\ref{fig:search}.  The mark vector $V$ in the node corresponding to the
last triplet `CDA' encodes the probable ending positions for strings with the
queried suffix.  The previous triplet, `BCD', is next considered.  Its position
vector is RIGHT SHIFT-ed by one position and is AND-ed with $V$, setting the
$2^{\text{nd}}$ and $3^{\text{rd}}$ bits of $V$.  The containers attached to the
last triplet `CDA' at these positions are finally searched to return the string
`ABCDA'.

Searching an unsuccessful suffix such as `ACDA' produces an empty $V$ vector as
there is no `ACD' triplet in the database.  Consequently, we directly report
that there are no strings with the queried suffix.  If the suffix `DCDA' is
queried, only the $3^{\text{rd}}$ bit of $V$ is set and the corresponding
container is searched.  Once more, this is an example of a false positive, as no
key with the queried suffix is found.

We now analyze the time complexity of this procedure.  In the worst case, every
mark bit is set and none of them gets pruned by the subsequent operations.  For
a suffix of length $f$, the complexity of performing the list operations is
$O(f.l)$, where $l$ is the maximum length of a key.  Finally, all $O(l)$
containers are searched.  Hence, the total time for suffix search is $O(fl) +
O(l) \times T$, where $T$ is the average time for searching a container.  

\subsection{Prefix Searching}
\label{sec:prefix}

The prefix searching method exploits the fact that a prefix of a key is a suffix
of the \emph{reverse} of the key.  Hence, we maintain a separate \sem{}
structure where the reverse of every key in the database is inserted.  A prefix
search in the original space translates to a suffix search on the reverse \sem{}
structure.  This strategy, however, doubles the space requirements of \sem{}.

\subsection{Substring Searching}
\label{sec:substring}

A substring can be efficiently searched in \sem{}, albeit with an increase in
the space requirements.  The key idea is to note that any substring, when
sufficiently shifted, becomes a prefix.  Thus, if the amount of shifting is
known, each key in the database can be shifted by that amount, and a prefix
search can be issued on the shifted keys.  This is precisely the idea that
\sem{} uses.

In addition to the original reverse \sem{} structure, we maintain $l-1$ extra
reverse structures, $S_i$, $i = 1, \dots, l-1$, where $l$ is the maximum
length of a key.  

\begin{figure}[t]
	\begin{center}
		\includegraphics[width=0.99\columnwidth]{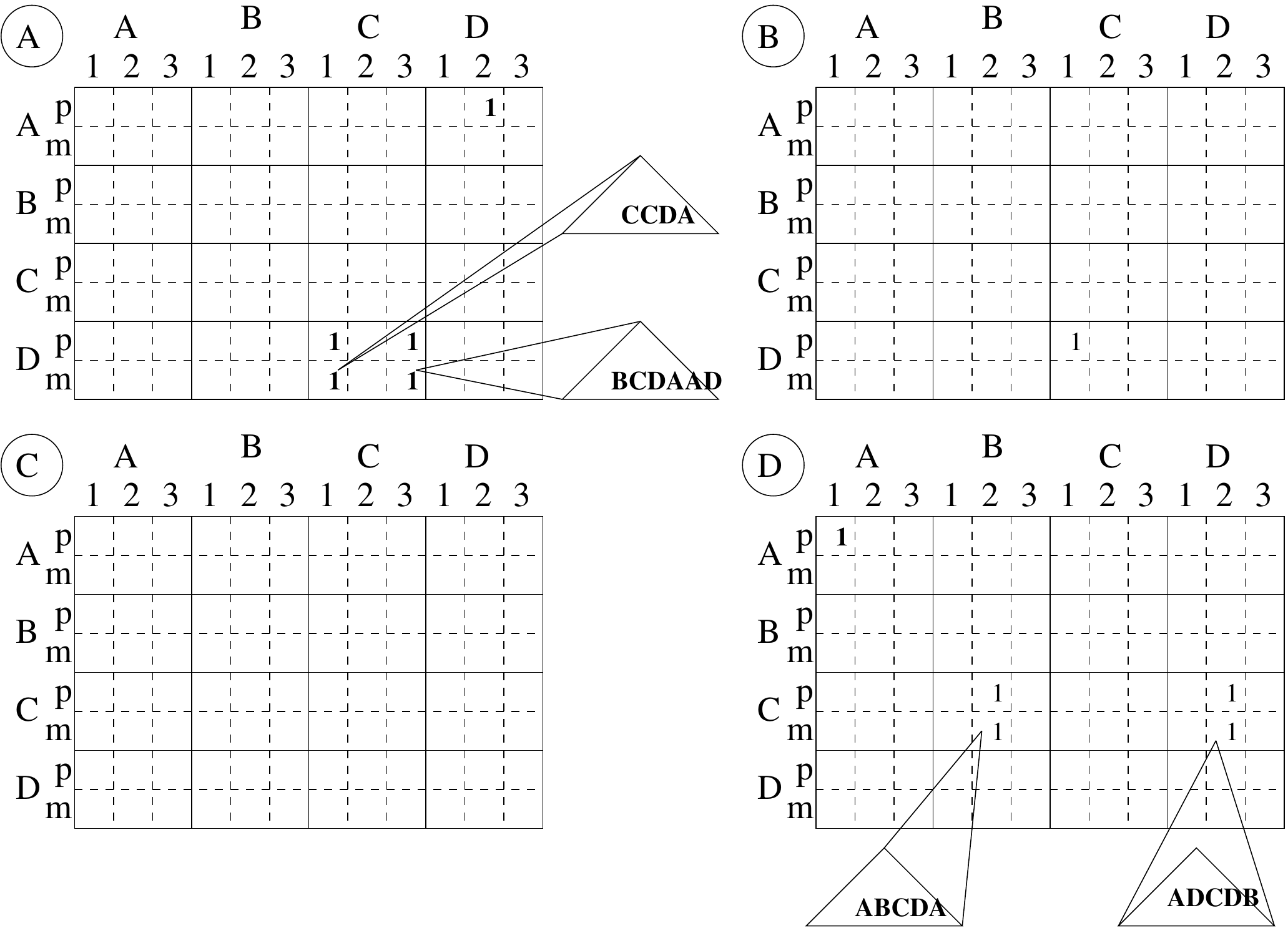}
		\caption{Example of extra reverse \sem{} structure for substring
		search.}
		\label{fig:substring}
	\end{center}
\end{figure}

Figure~\ref{fig:substring} shows the first reverse structure $S_1$ corresponding
to the keys in Figure~\ref{fig:search}.  When a key of length $n$ is inserted
into \sem{} and its reverse is inserted into reverse \sem{}, $n-1$ strings are
extracted from the key in addition, by shifting one character at a time.  The
resulting strings are inserted into the corresponding reverse \sem{} structures.
Suppose a key $K = c_1c_2 \dots c_n$ is inserted.  We create $n-1$ strings from
the key.  The $i^\text{th}$ string $K_i = c_{i+1}c_{i+2} \dots c_n$ is inserted
into $S_i$.  Although only a part of the key, i.e., $K_i$ is used to index in
$S_i$, the containers of $S_i$ stores the entire key $K$.  This is done to
ensure that the original keys can be returned from $S_i$ after a successful
search.

The algorithm for substring search uses a similar strategy as the suffix search.
When a substring $c_1c_2 \dots c_r$ of length $r$ is queried, first, the
positions where the last triplet $c_{r-2}c_{r-1}c_r$ are present are found by
using the position array corresponding to the triplet.  Note that this deviates
from the suffix search as the position vector, and not the mark vector needs to
be searched, since a key may not necessarily end with the substring.  The
triplets are then traversed backwards and all possible positions where the
substring can start are found.  Suppose the list of these positions is $L$.  For
every position $p \in L$, a prefix search with the substring is performed at the
structure $S_p$.  The complete results on searching the various structures
provides the entire set of database keys containing the substring.

Consider a substring query for `BCDA' in the \sem{} structure shown in
Figure~\ref{fig:search}.  The position bit vector $V$ of the last triplet `CDA'
includes all possible positions where the substring can end in a key.  Next, $V$
is then LEFT SHIFT-ed by one bit and bit-wise AND-ed with the position vector of
the previous substring triplet, i.e., `BCD'.  The bits at position $1$ and $2$
of $V$ are set in this process.  This implies that the substring can start only
at positions $1$ and $2$ in a key.  Hence, a prefix search with `BCDA' is issued
in the two reverse \sem{} structures corresponding to these positions, i.e., the
(original) reverse \sem{} and the one-shifted reverse \sem{} $S_1$.  The prefix
searches in the two structures generate `ABCDA' and `BCDAAD' as the result.

Next, consider an unsuccessful substring search on `CCDD'.  Since there is no
such triplet `CDD' in the database, the search can be immediately terminated
without accessing any of the reverse structures.  This provides a substantial
advantage over other brute-force or trie-based methods.  

The chances of false positives, however, remain.  For example, consider the
substring `DCDA'.  The position vectors of `CDA' when LEFT SHIFT-ed and AND-ed
with the position vector of `DCD' yield position $2$ as a possibility where the
substring can occur in a key.  Thus, a prefix search in the one-shifted reverse
structure $S_1$ is issued.  However, only an empty result set is returned.

Storing the extra \sem{} structures increases the total space complexity to
$2k^3l^2$ bits.  For the English dictionary mentioned in
Section~\ref{sec:structure}, this evaluates to 3.5\,MB.  If there is not enough
space in the memory to store all the reverse structures, the extra ones are
stored on disk.  These extra structures are invoked only for a substring search,
and only if the corresponding offset is in the possible list of positions.  As
the extra \sem{} structures are independent, the prefix searches in the
different structures can be performed in parallel.  The experiments reported in
Section~\ref{sec:expts}, however, do not use parallelization.

In a sequential machine, the time for substring search is determined by the
number of prefix searches and the time for each of them.  So, the total time
complexity for searching a substring of length $r$ is $(O(lr) + O(l)\times T)
\times$ (the number of prefix search positions found), where $T$ is the average
time to search a container.  In the next section, we calculate the expected
number of such prefix searches.

\subsection{Analysis of Prefix, Suffix, and Substring Searching}

The search procedures guarantee correct results by finally searching the
containers that have a possibility of containing an answer.  An unsuccessful
search may be generated if all the triplets present in the query are also
present at the same position in other keys of the database.  We now analyze the
searching of suffixes, prefixes and substrings.

Eq.~\eqref{eq:entire_bound} shows the probability that a particular string of
length $n$ is searched in a container.  The probability $P_{prefix}$ that the
entire prefix of length $s$ is matched, and an unsuccessful search is generated,
can be deduced similarly:
\begin{align}
	\label{eq:prefix}
	P_{prefix} &\leq 1-3(s-2)\left(1-1/k\right)^{m}
\end{align}
Note that here we are ignoring the positions where a prefix can start as we have
bounded the number of keys at a position by its worst case, which is the total
number of keys $m$.  In reality, $P_{prefix}$ is much less.  The probability
$P_{suffix}$ that the entire length of a suffix of length $f$ is matched, and an
unsuccessful search is generated is the same when $f(i)$ and $g(i)$ terms are
bounded by $m$.

The above equation also provides an upper bound of the probability that a search
for a substring of length $s$ is issued when it is not present in the database.
We use this bound to analyze the substring searching.

The substring search is actually a series of prefix searches.  Each such search
has an analysis as given by Eq.~\eqref{eq:prefix}.  The expected number of
prefix searches that will be issued in the different \sem{} structures for a
substring search is equal to the expected number of positions in the final list
after all the triplets of the substring have been traversed.

We assume the event that the substring of length $s$ occurs at position $i$ to
be independent of the event that the substring occurs at some other position
$j$.  Again, this is a simplification, as for long substrings or for short
differences in $i$ and $j$, the events are not independent.  Modeling the
occurrence of the substring by binomial trials, the expected number of positions
where the substring occurs is given by the product of the total number of trials
and the probability of success in each trial.  The total number of trials is $l$
as there can be $l$ positions.  The probability of success in each trial (i.e.,
position), is given by Eq.~\eqref{eq:prefix}.  The expected number of prefix
searches is then
\begin{align}
	\label{eq:expected}
	l \times P_{prefix} \leq l \times \left( 1-3(s-2)\left(1-1/k\right)^{m} \right)
\end{align}

When the largest length of a key, $l$, increases, the chance that a prefix
search needs to be issued also increases.  When the number of keys, $m$,
increases, it becomes more likely that a key in the database will have the
queried substring, thereby increasing the number of searches.  The length of the
substring queried, $s$, has an opposing effect as more triplets need to be
present before a search is issued in the container.  Finally, when the size of
the alphabet, $k$, increases, the chance that a particular triplet occurs
decreases since the probability of a character matching with another decreases.

\subsection{Deletion, Updating, and Re-insertion}

When a key is to be deleted from \sem{}, it is first searched.  If it is found,
the deletion operation in the container is performed.  The corresponding mark
bit is reset to $0$ only when the container becomes empty; also, the container
is de-allocated.  Updating a key involves deleting the key and then inserting
the modified key, while re-insertion follows the same procedure as insertion.
The time complexities of these procedures are bounded by those of insertion and
searching.  

The mark vector and the position vectors remain filled up after repeated
deletions and insertions.  This poses a problem for searching as the pruning
capacity of \sem{} decreases.  However, unlike the position bits, a mark bit can
be reset to $0$ if the corresponding container becomes empty due to deletions.
In any case, the time for searching in the container decreases even though the
mark bit remains set (if the container does not become empty).  Further, most
string-based applications perform many more insertion and search operations than
deletion, thereby rendering this a not-so-critical issue.

\section{Experiments}
\label{sec:expts}

In order to assess the performance of \sem{}, we conducted tests on multiple
datasets and compared it with two other structures, burst trie~\cite{burst-7}
and compact trie~\cite{trie-32,tree-20}.  While there exists a number of other
structures that support string operations (see Section~\ref{sec:survey}), the
burst trie is reported to require the least amount of memory~\cite{hat-8}, while
the compact trie is reported to be the fastest for exact key searching
operations~\cite{trie-32,tree-20}.  Hence, we compared \sem{} with these two
structures only.  

We used two real datasets: (i)~English dictionary (obtained from
\url{http://www.outpost9.com/files/WordLists.html}), and (ii)~protein sequences
from RCSB Protein Data Bank (PDB, \url{ http://www.rcsb.org/pdb/}).  We also
used synthetic datasets to assess the scalability and practicality of our
algorithms.  The datasets were uniformly distributed random data (henceforth
referred to as Uniform dataset) and Zipfian distributed data (Zipfian dataset),
both with varying parameters.  Section~\ref{sec:anal_searching} assumes a random
distribution while many natural datasets such as the English dictionary follow
the Zipfian distribution.

The containers in \sem{} can be organized as a list or as a BST.  These two
variants were compared against the two trie variants, burst trie and compact
trie, with respect to the following parameters: (i)~memory size, (ii)~insertion
time, and (iii)~searching time for both successful and unsuccessful searches.
We also measure empirically the probability of pruning the false positives
during a search as well as show the results for prefix, suffix, and substring
searches.  These experiments were run on a 2.1\,GHz desktop PC with 2\,GB of
memory using C++ compiler on a Linux platform.  Due to space constraints, we
show only the representative results while complete results can be found
in~\cite{thesis}.

\subsection{Real datasets}

\begin{table*}[t]
	\begin{center}
		\begin{small}
			\begin{tabular}{|c||c|c|c|c|c|c|c|}
				\hline
				\multirow{2}{*}{Dataset} & Number of & Number of & 
				Longest key & Number of & Max. size of & 
				Avg. size of & False positive \\
				& keys, $m$ & symbols, $k$ & length, $l$ & 
				characters & a container & a container & rate \\
				\hline
				\hline
				English dictionary & 179,935 & 26 & 45 & 1,198,635 & 601 & 7.5 & 0.019 \\
				\hline
				Protein sequences & 38,627 & 21 & 2512 & 5,846,331 & 205 & 1.3 & 0.161 \\
				\hline
			\end{tabular}
		\caption{Parameters and search performance for the real datasets.}
		\label{tab:data}
		\end{small}
	\end{center}
\end{table*}

\begin{table*}[t]
	\begin{center}
		\begin{small}
		\begin{tabular}{cc}
			\hspace*{-2mm}
			\begin{tabular}{|c||c|c|c|c|c|}
				\hline
				Index & Total & Time to & \multicolumn{3}{c|}{Searching time} \\
				\cline{4-6}
				structure & memory & insert & Succ & Unsucc & Total \\
				\hline
				\hline
				INS. BST & 1.50\,MB & 1.42\,s & 0.51\,s & 0.54\,s & 1.05\,s \\
				\hline
				INS. List & 1.50\,MB & 1.29\,s & 0.59\,s & 0.58\,s & 1.17\,s \\
				\hline
				Burst tr. & 1.53\,MB & 1.61\,s & 0.64\,s & 0.66\,s & 1.30\,s \\
				\hline
				Compact tr. & 2.38\,MB & 1.82\,s & 0.65\,s & 0.65\,s & 1.31\,s \\
				\hline
			\end{tabular}
			& \hspace*{-4mm}
			\begin{tabular}{|c||c|c|c|c|c|}
				\hline
				Index & Total & Time to & \multicolumn{3}{c|}{Searching time} \\
				\cline{4-6}
				structure & memory & insert & Succ & Unsucc & Total \\
				\hline
				\hline
				INS. BST & 15.73\,MB & 4.89\,s & 2.28\,s & 2.21\,s & 4.49\,s \\
				\hline
				INS. List & 15.73\,MB & 4.66\,s & 2.44\,s & 2.16\,s & 4.60\,s \\
				\hline
				Burst tr. & 15.89\,MB & 5.64\,s & 2.64\,s & 2.67\,s & 5.31\,s \\
				\hline
				Compact tr. & 25.71\,MB & 9.29\,s & 2.70\,s & 2.37\,s & 5.07\,s \\
				\hline
			\end{tabular} \\
		(a) & (b)
		\end{tabular}
		\caption{(a) English dictionary results. (b) Protein sequence results.}
		\label{tab:res}
		\end{small}
	\end{center}
\end{table*}

Table~\ref{tab:data} summarizes the two real datasets.  Table~\ref{tab:res}(a)
shows that the \sem{} structures require lesser storage space than the other two
structures.  The main component of the storage comes from the actual keys
themselves, and thus, the differences are very small.  The insertion and search
times are also better.  Table~\ref{tab:res}(b), on the other hand, shows that
the memory requirement of the \sem{} structure becomes very large when the
length of the keys are large.  The overhead of maintaining bit vectors of length
$2512$ for every cell of the matrix requires about 10\,MB of memory space.
However, the insertion and search times are lesser than those for the burst
trie.  The pruning offered by indexing makes the search faster.

Since the search performance of \sem{} depends on the number and size of
containers, we measured the following additional parameters as well: (i)~total
number of containers, (ii)~largest size of a container, and (iii)~average size
of a container.

The average size of a container shows how well the keys are spread.  If this
number is low, then the keys are well-distributed in the containers.  Then, even
when a container is accessed for a key that is absent in the database, the
overhead of searching the container is less.  In such cases, the choice of the
list versus BST variants does not matter much.

The other important factor for searching time is the false positive rate.  It is
measured as the number of times a container is accessed and searched for a
search key that is not in the database, i.e., for an unsuccessful search.
Table~\ref{tab:data} shows that this ratio is almost negligible for the
dictionary dataset.  Thus, the index in the \sem{} structure can prune
efficiently most of the unsuccessful searches without accessing the containers.
Even for the protein dataset, about 84\% of the unsuccessful searches are
pruned.

\subsection{Uniform and Zipfian datasets}

For synthetic datasets, the important parameters affecting the performance of
the algorithms are: (i)~total number of keys, $m$, (ii)~size of the alphabet,
$k$, (iii)~length of the longest key, $l$, and (iv)~length of the query
substring, $n$.

The datasets were generated by controlling these parameters.  The length of each
key was chosen randomly from $1$ to $l$, and each character was chosen from an
uniform or a Zipfian distribution of $k$ characters.  Two-thirds of the keys
thus generated were inserted in the structure.  The rest one-third was used to
trigger searches that were unsuccessful.  Half of the inserted data (i.e.,
one-third of the total generated keys) was used to trigger successful searches.
The prefix, suffix and substring were generated from the strings stored,
starting from random positions and of varying lengths.

\subsection{Effect of number of keys}

\begin{figure}[t]
	\begin{center}
		\begin{tabular}{cc}
		\hspace*{-3mm}
		\includegraphics[width=\figwidth]{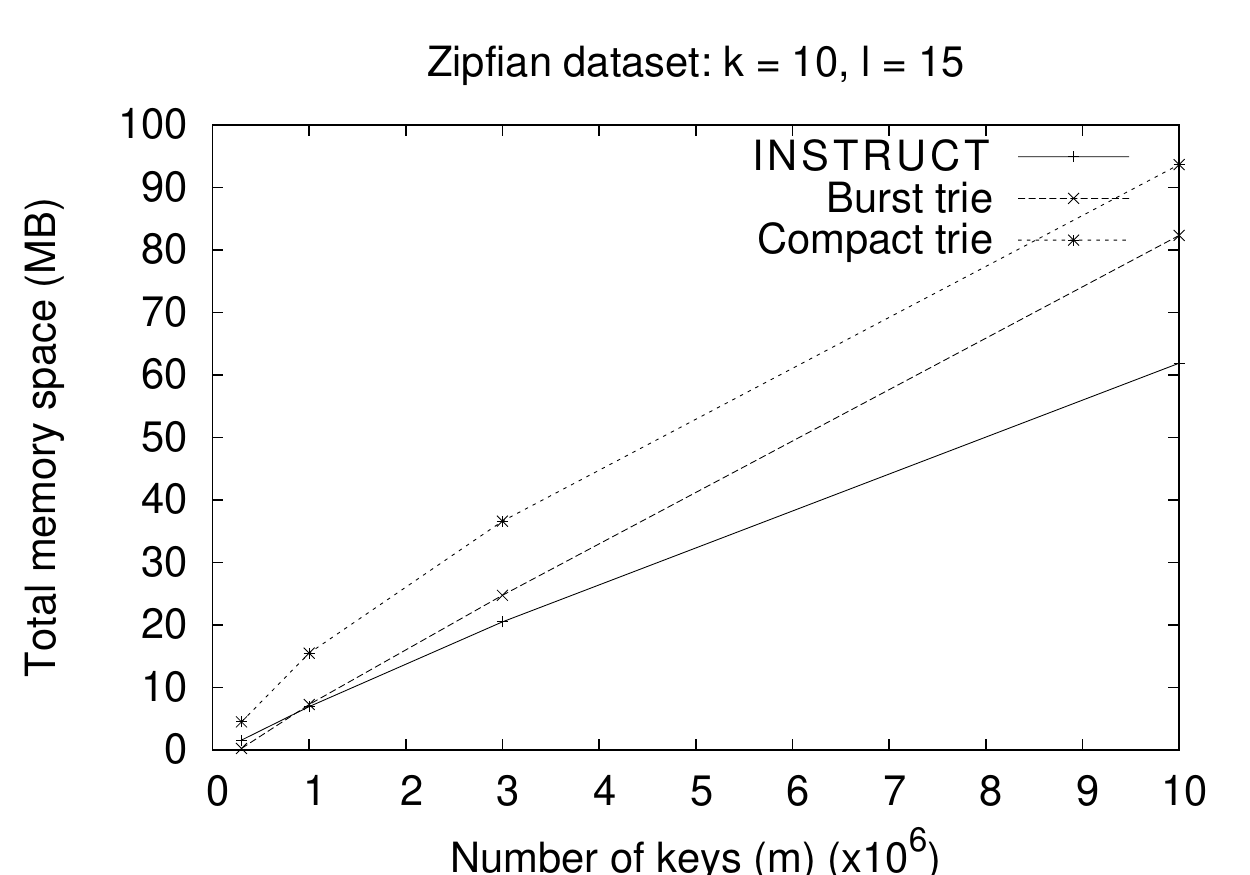} &
		\hspace*{-6mm}
		\includegraphics[width=\figwidth]{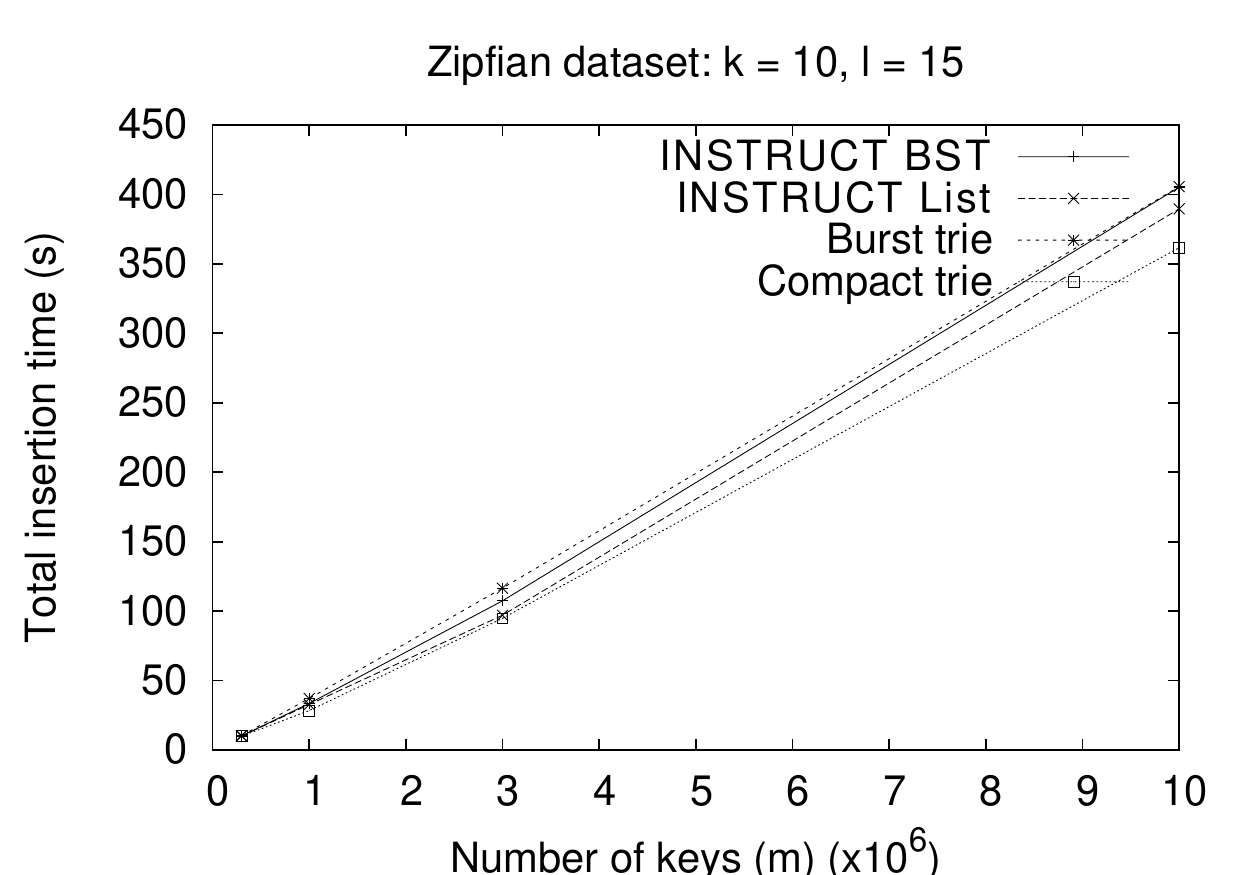} \\
		(a) & (b)
		\end{tabular}
		\caption{Effect of number of keys on (a) memory size and (b) insertion time.}
		\label{fig:zip_m}
	\end{center}
\end{figure}

With the increase in the number of keys, the size of dataset increases.
Therefore, the memory requirement increases as well.  However, the size of the
multi-dimensional array index structure of \sem{} is independent of the number
of keys.  It depends only on the alphabet set size and the length of keys.
Hence, the growth in memory space is at most linear due to the actual key
storage in the containers.  Figure~\ref{fig:zip_m}(a) shows that \sem{} requires
the least amount of memory and has a better scalability as compared to the burst
and compact tries.

Figure~\ref{fig:zip_m}(b) shows the effect of number of keys on the insertion
time for Zipfian data.  As expected, the scalability is roughly linear for all
the structures.  As the number of keys increases, the average size of each
container increases as well.  This explains the widening gap in insertion times
between the two variants of \sem{}.  The burst trie performs the worst due to
the nature of the burst heuristic.

\begin{figure}[t]
	\begin{center}
		\begin{tabular}{cc}
		\hspace*{-3mm}
		\includegraphics[width=\figwidth]{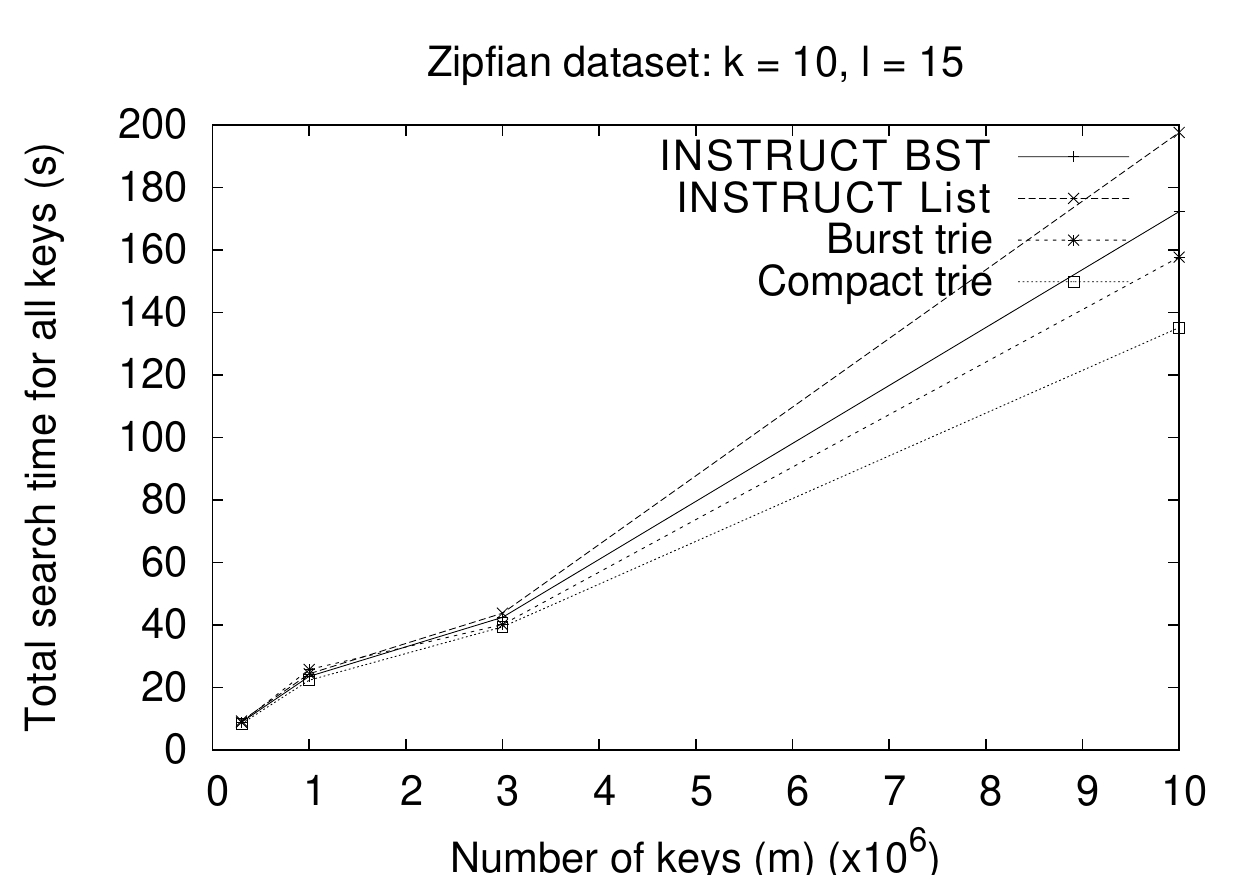} &
		\hspace*{-6mm}
		\includegraphics[width=\figwidth]{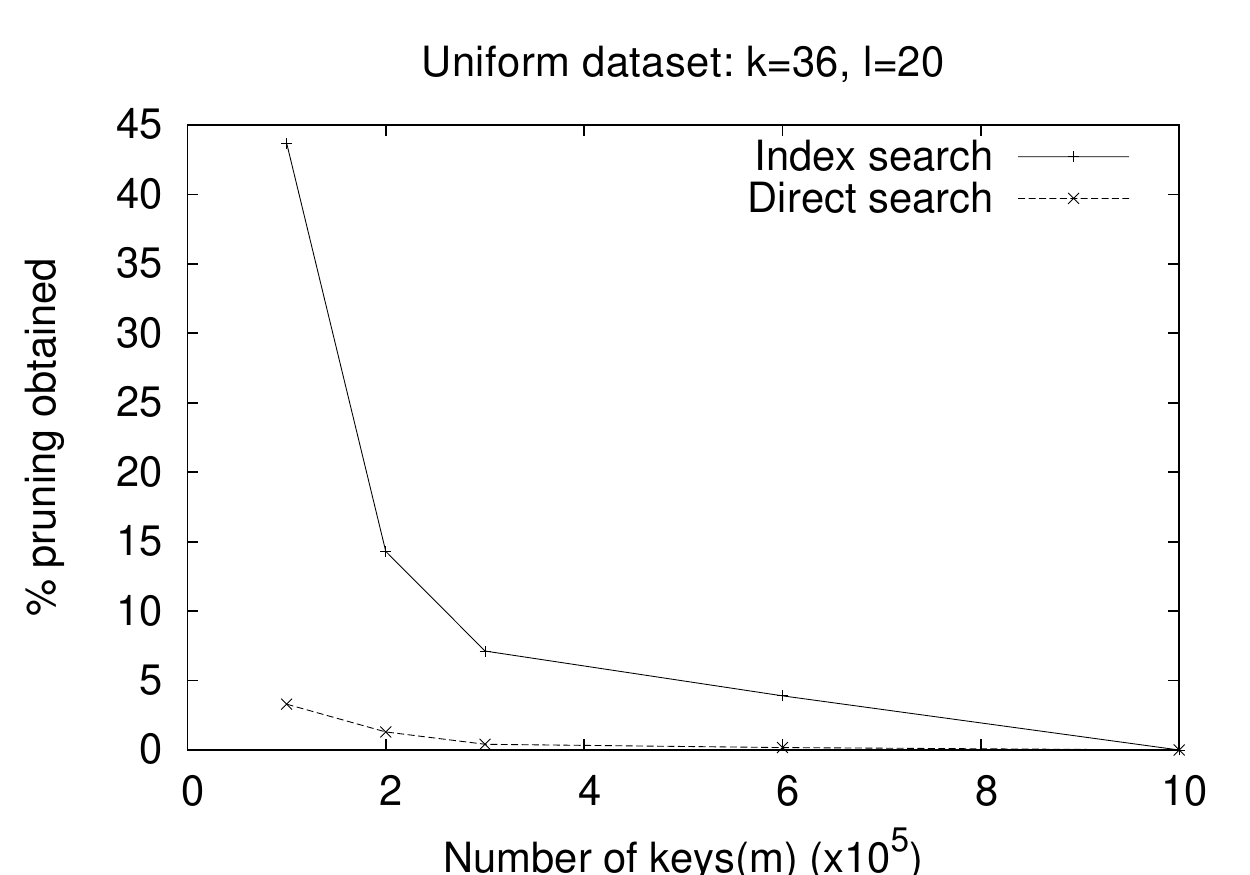} \\
		(a) & (b)
		\end{tabular}
		\caption{Effect of number of keys on (a) search time and (b) pruning.}
		\label{fig:prn_src_m}
	\end{center}
\end{figure}

The next experiment measures the running time for searching both successful and
unsuccessful keys.  The performance of \sem{} suffers when a large number of
keys are present in the database (Figure~\ref{fig:prn_src_m}(a)).  The large
number of false positives with the increase in the size of the database
necessitates more searches in the containers.  The large size of the containers
degrades the search performance.  The BST variant performs better than the list
variant due to its superior arrangement of keys in the container.  Modeling the
list in a lexicographic order would help in boosting the performance of the 
list implementation of the containers.

To analyze the search time for unsuccessful keys of the direct search strategy
versus the index search strategy, we measured the ratio of the number of
searches pruned.  Figure~\ref{fig:prn_src_m}(b) shows the comparison of the
ratio of pruning between the two strategies.  The pruning for the direct
strategy is almost constant while that for the index strategy decreases
exponentially with the number of keys as indicated by
Eq.~\eqref{eq:entire_bound}.  The figure also illustrates the fact that it is
prudent to follow the direct search when there is a large number of keys as it
is more likely that all the triplets checked will be in the database and the
search cannot be pruned (as the pruning factor for both the strategies roughly
becomes the same), thereby reducing the actual search time.

\subsection{Effect of largest key length}

\begin{figure}[t]
	\begin{center}
		\begin{tabular}{cc}
		\hspace*{-3mm}
		\includegraphics[width=\figwidth]{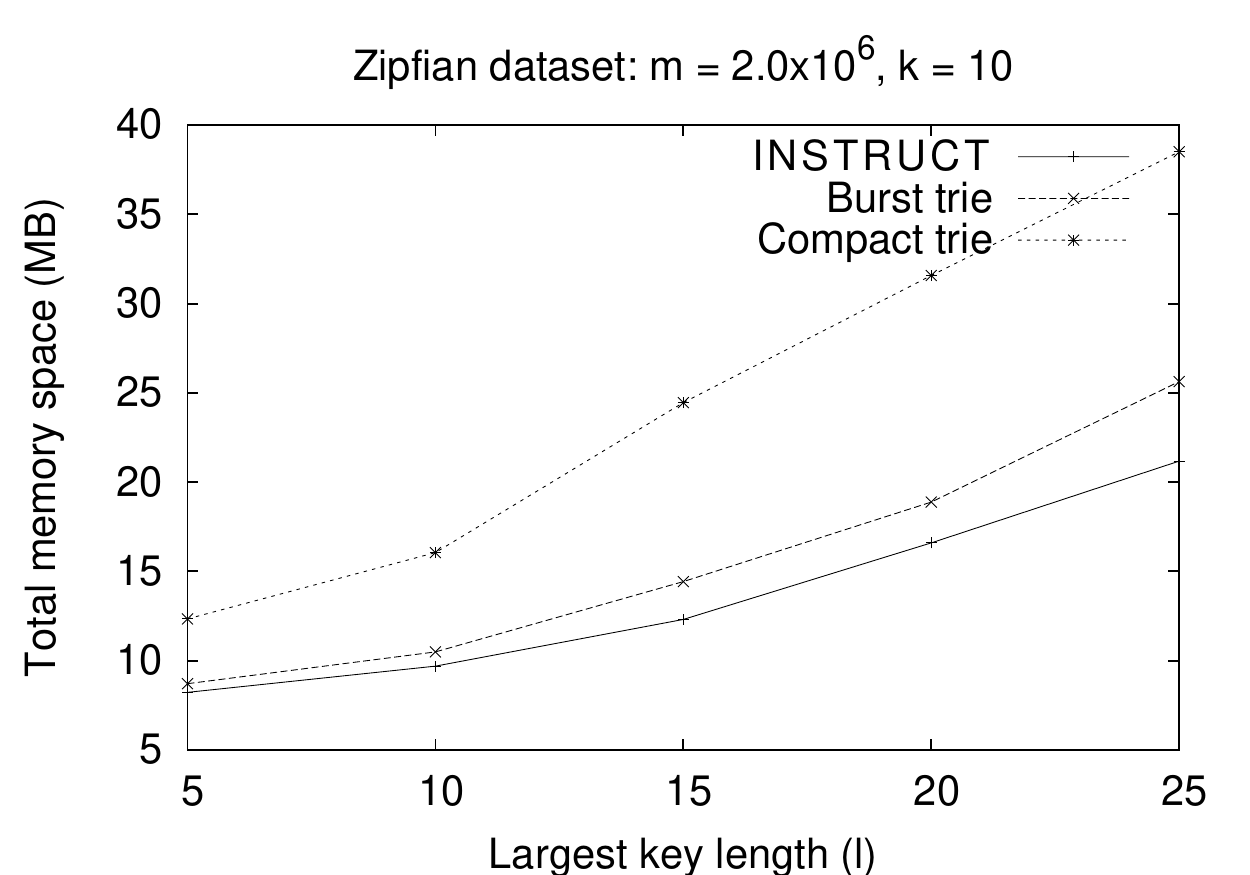} &
		\hspace*{-6mm}
		\includegraphics[width=\figwidth]{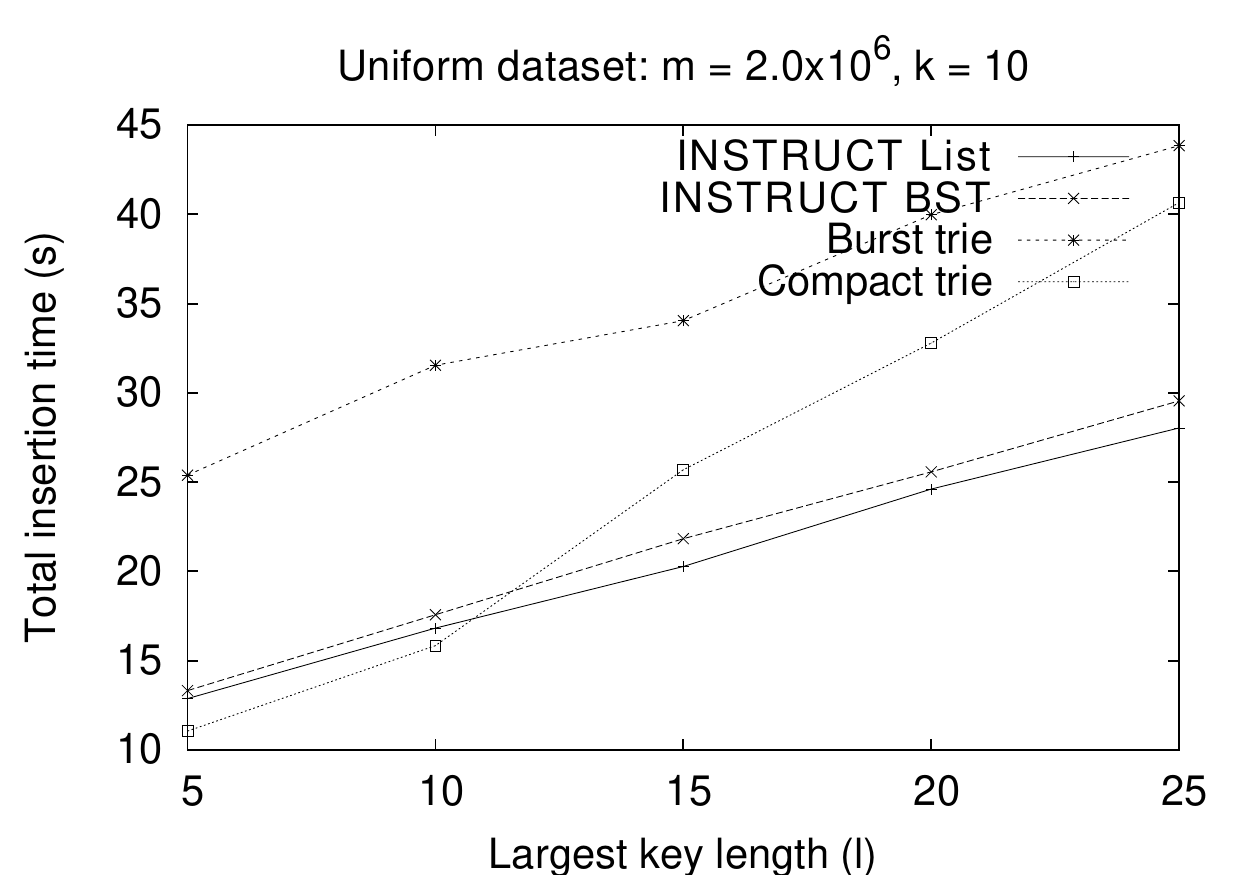} \\
		(a) & (b)
		\end{tabular}
		\caption{Effect of largest key length on (a) memory size and (b) insertion time.}
		\label{fig:zip_l}
	\end{center}
\end{figure}

The next set of experiments measure the effect of the key length on the various
algorithms.  The number of pointers in trie-based structures increases with the
maximum length of the keys.  Figure~\ref{fig:zip_l}(a) shows that the increase
in memory size with the largest key length is faster for the trie-based
structures.  In the case of \sem{}, only the lengths of the bit vectors increase
and, thus, the size of the whole index increases linearly.  However, the memory
requirement is mainly dominated by the actual storage of the keys, and
therefore, the scalability is much better.  Consequently, \sem{} requires lesser
memory space (refer~\cite{thesis}).

\begin{figure}[t]
	\begin{center}
		\begin{tabular}{cc}
		\hspace*{-3mm}
		\includegraphics[width=\figwidth]{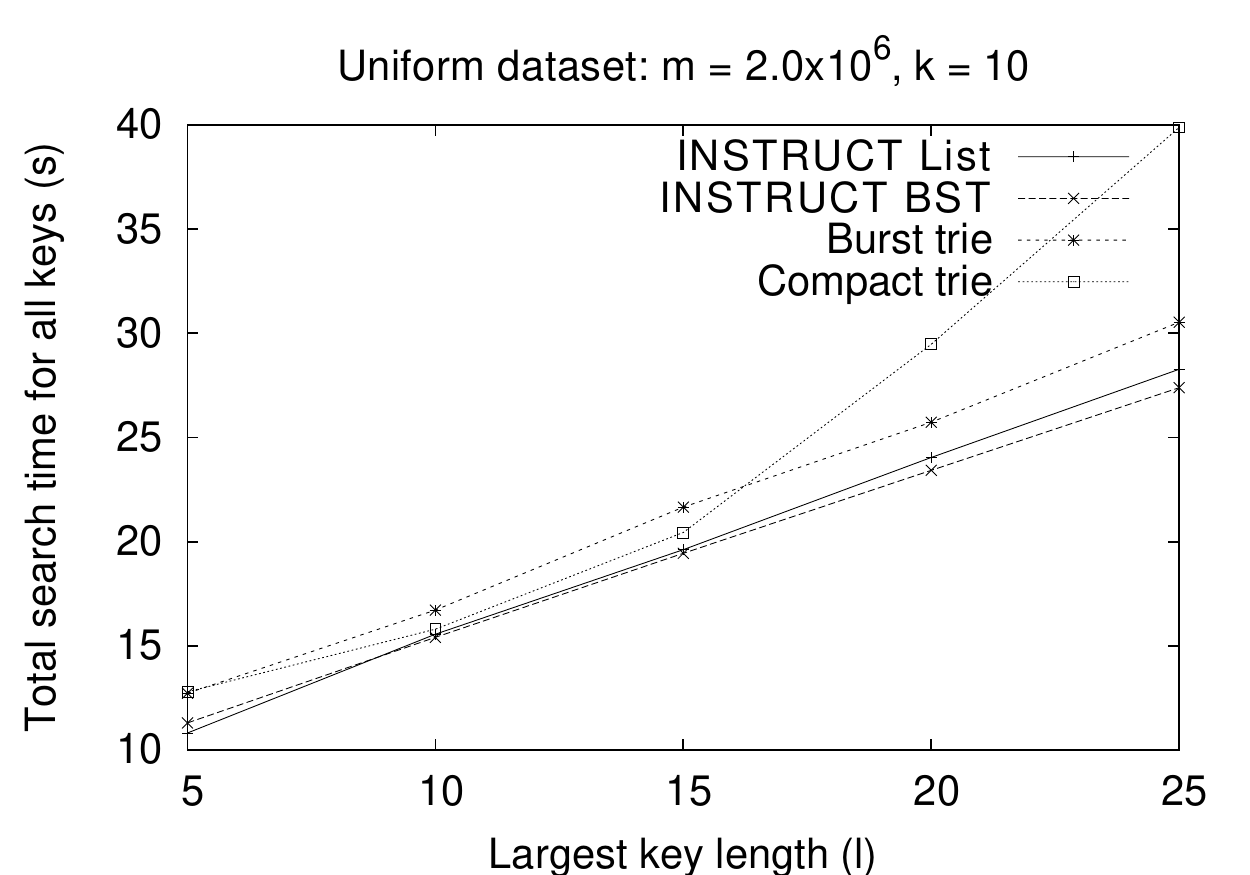} &
		\hspace*{-6mm}
		\includegraphics[width=\figwidth]{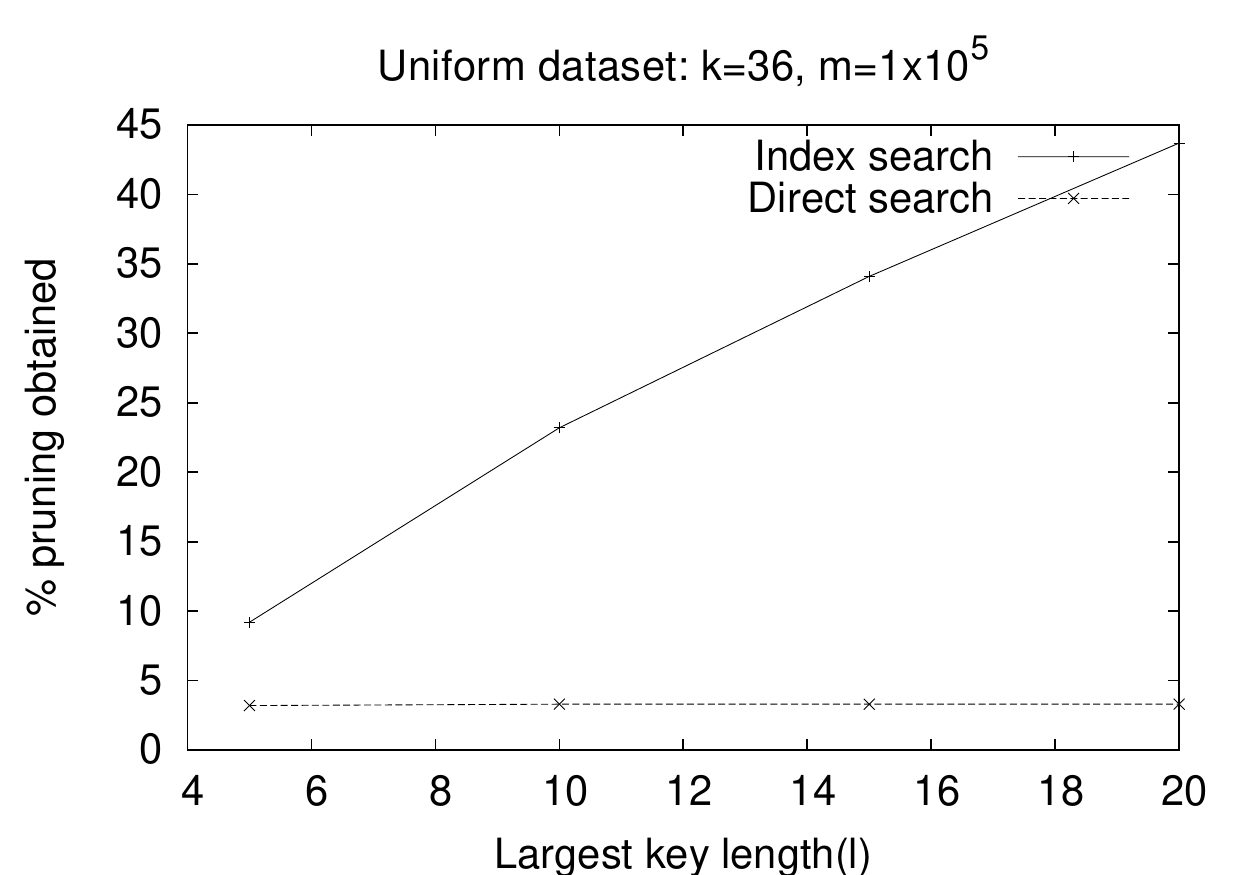} \\
		(a) & (b)
		\end{tabular}
		\caption{Effect of largest key length on (a) search time and (b) pruning.}
		\label{fig:ser_l}
	\end{center}
\end{figure}

Inserting a key requires setting the bits corresponding to all the triplets in
the index; so, the insertion time increases with the key length
(Figure~\ref{fig:zip_l}(b)).  However, since trie-based structures invoke
pointer chasing whereas \sem{} uses direct array access, the insertion procedure
in \sem{} is faster.

Searching a key with a larger length has two opposing effects on the running
time.  On one hand, more number of triplets need to be checked in the structure.
On the other hand, Eq.~\eqref{eq:entire} shows that more the number of triplets
in a key, the better is the chance of pruning it, thereby saving the searching
time inside a container.  However, for successful searches, the time to search
in the index is simply an overhead, as the container will have to be searched.
Thus, the total time for searching increases.  Nevertheless, the searching times
using \sem{} are smaller than the trie structures (Figure~\ref{fig:ser_l}(a)).

Figure~\ref{fig:ser_l}(b) shows that the pruning produced by larger number of
triplets in a longer key makes searching through the index perform better than
the direct search.  The increase in pruning is linear with the length of the
key, as expected from Eq.~\eqref{eq:entire_bound}, making the indexed strategy
better for longer keys.

\subsection{Effect of alphabet size}

\begin{figure}[t]
	\begin{center}
		\begin{tabular}{cc}
		\hspace*{-3mm}
		\includegraphics[width=\figwidth]{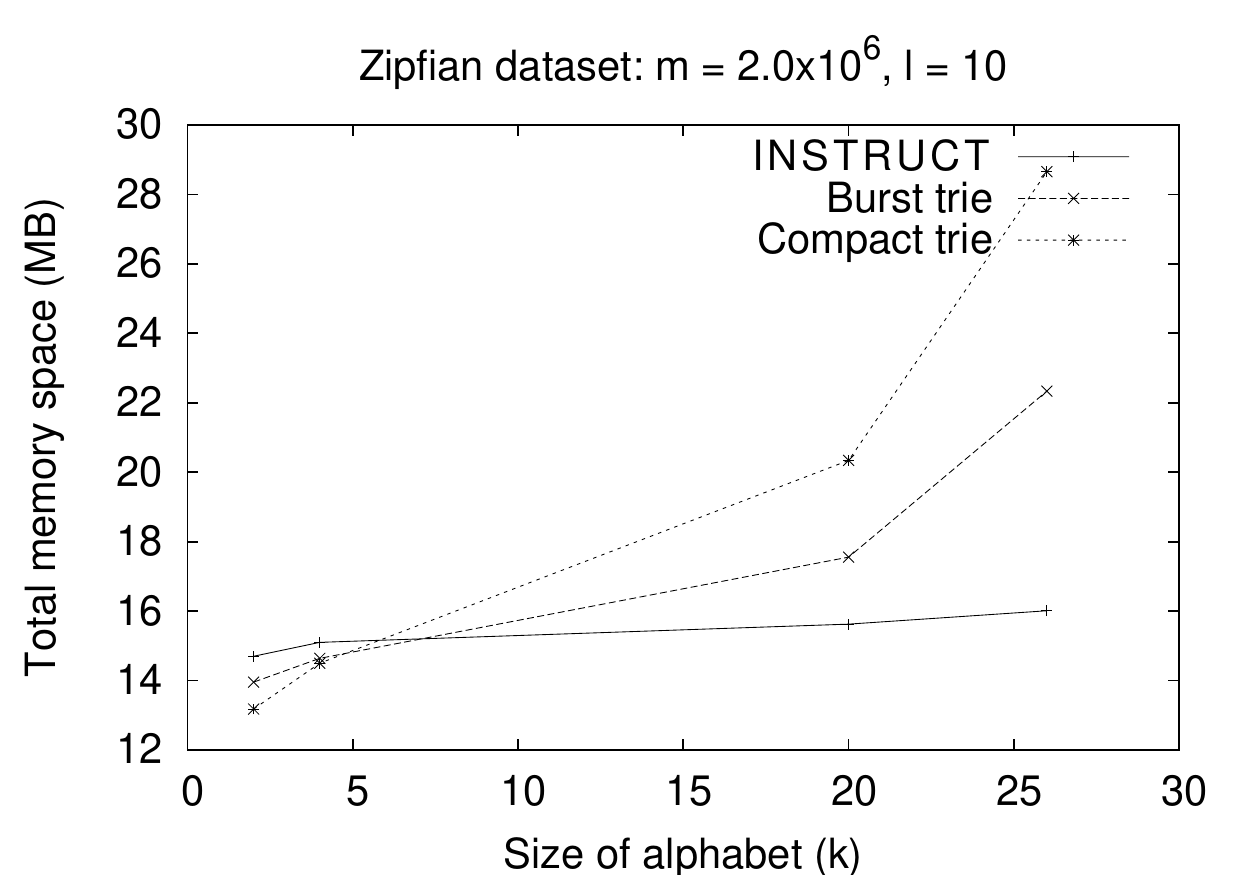} &
		\hspace*{-6mm}
		\includegraphics[width=\figwidth]{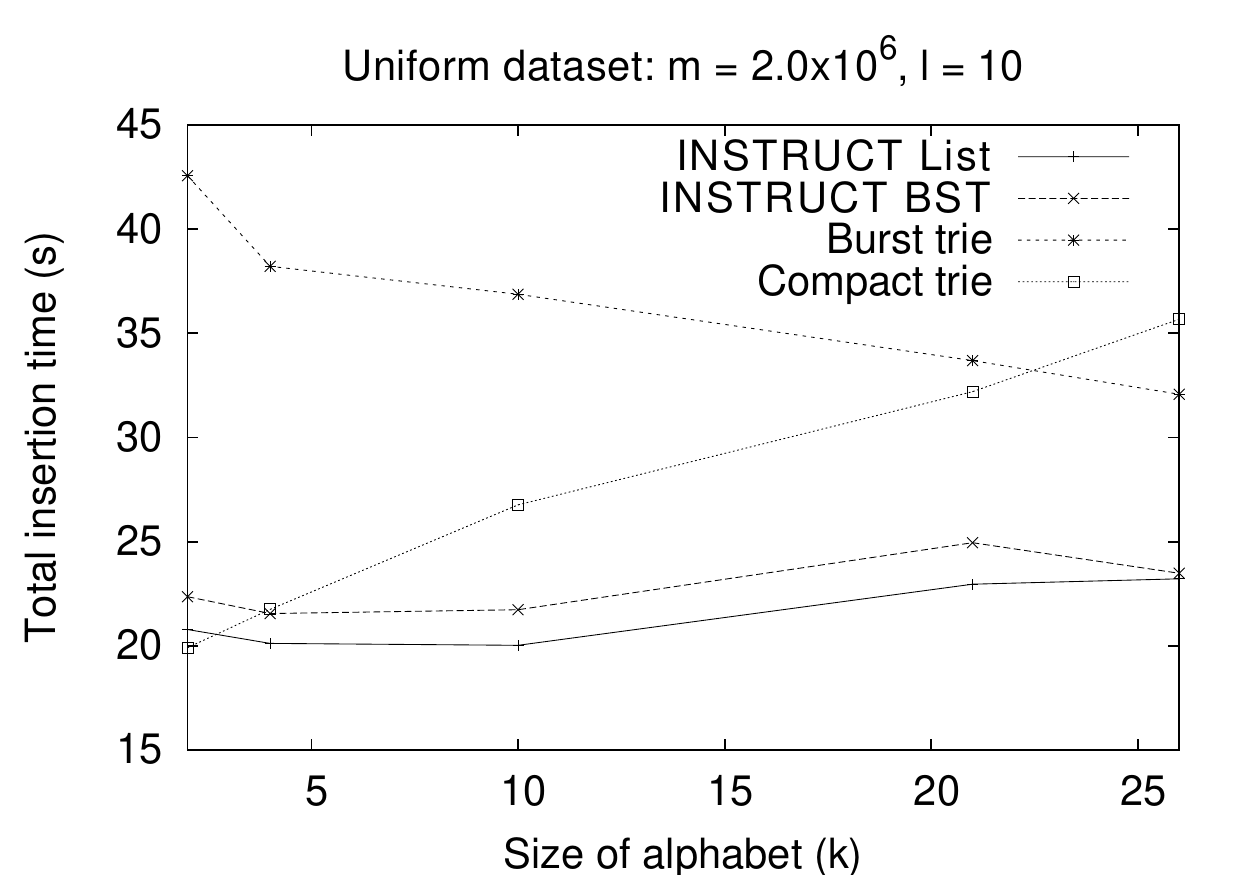} \\
		(a) & (b)
		\end{tabular}
		\caption{Effect of alphabet size on (a) memory size and (b) insertion time.}
		\label{fig:zip_k}
	\end{center}
\end{figure}

\begin{figure}[t]
	\begin{center}
		\begin{tabular}{cc}
		\hspace*{-3mm}
		\includegraphics[width=\figwidth]{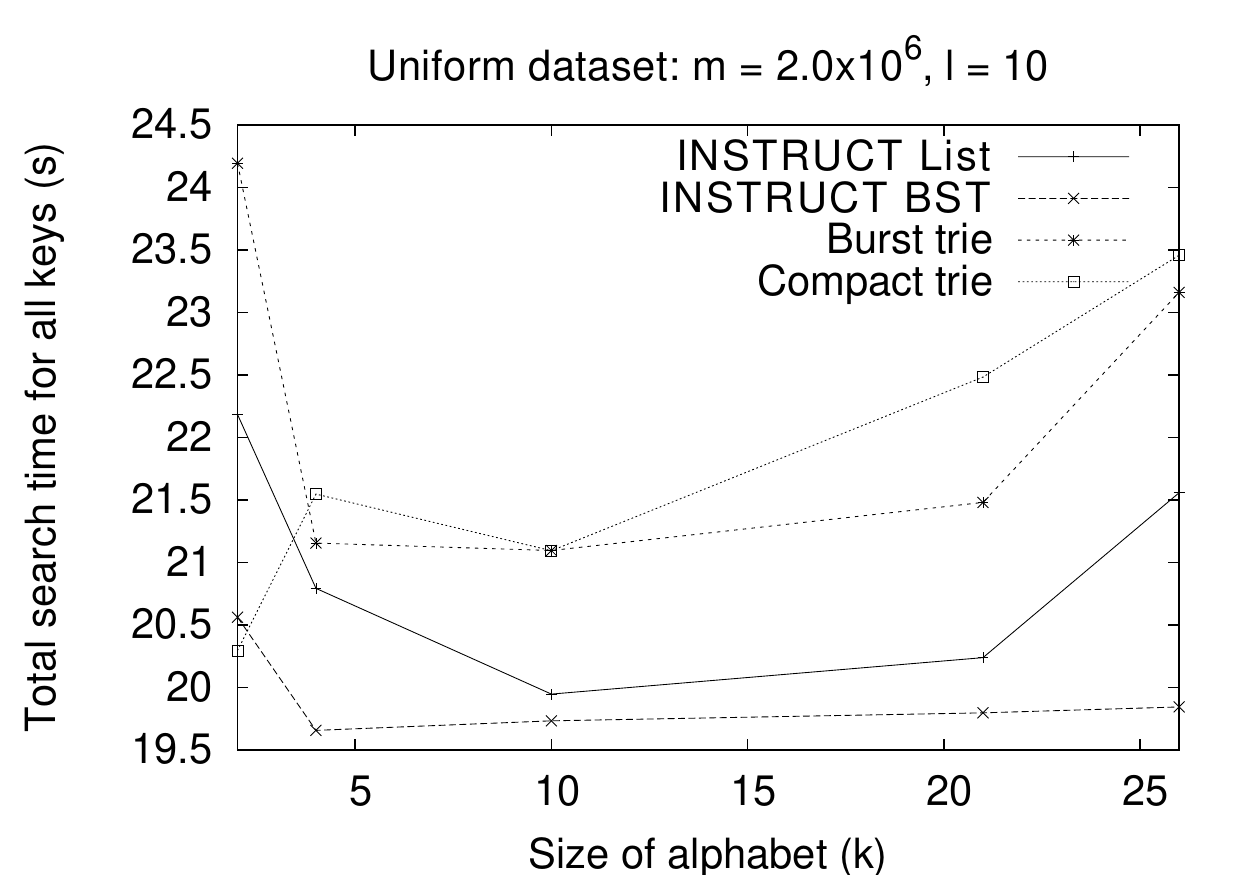} &
		\hspace*{-6mm}
		\includegraphics[width=\figwidth]{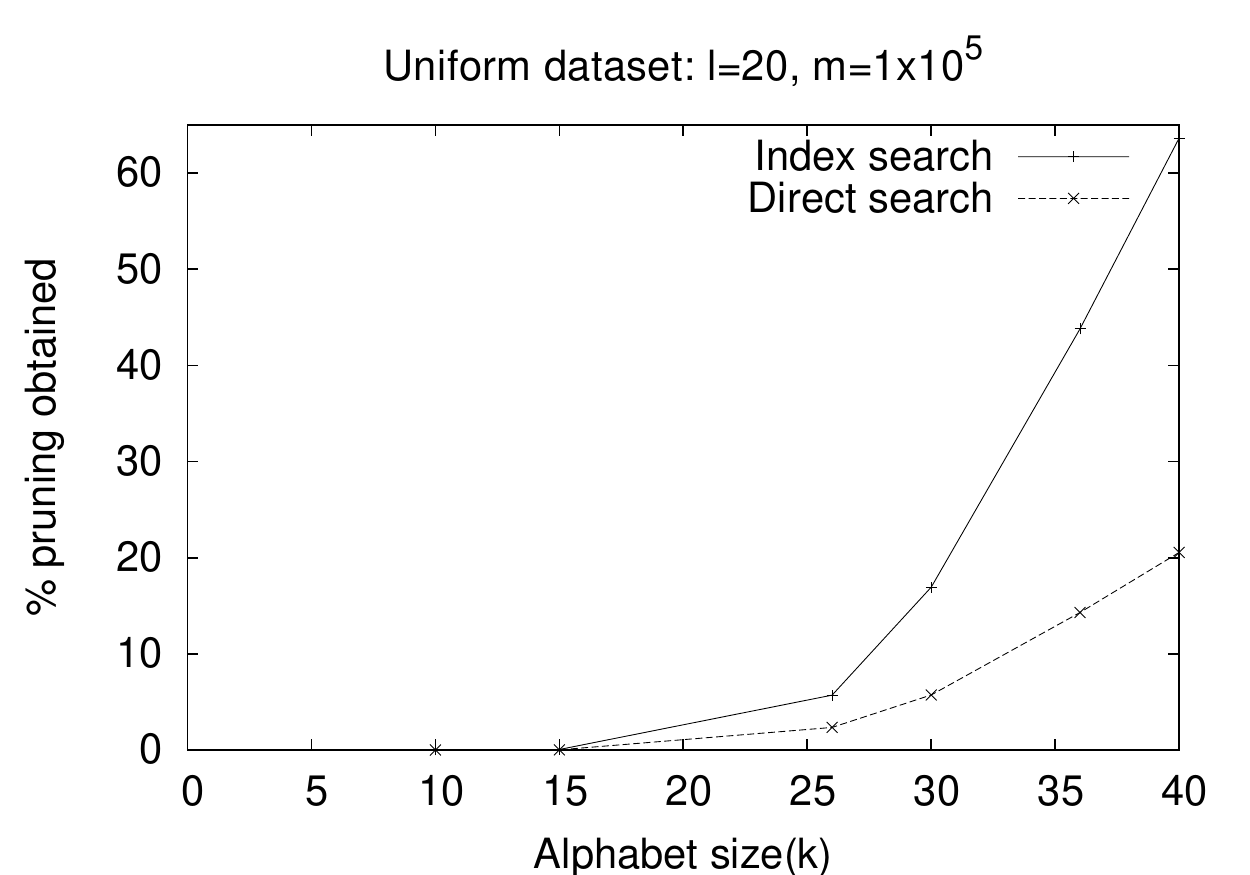} \\
		(a) & (b)
		\end{tabular}
		\caption{Effect of alphabet size on (a) search time and (b) pruning.}
		\label{fig:ser_k}
	\end{center}
\end{figure}

With the increase in the number of characters, the fanout of the trie-based
structures increases.  Due to this increase in the number of pointers, the
memory requirement increases (Figure~\ref{fig:zip_k}(a)).  In \sem{}, even
though the size of the index increases cubically, it is only in the order of
bits.  Thus, the size of the memory increases only slightly.

For a larger alphabet size, the spread of the keys becomes better due to lesser
number of collisions.  Consequently, the burst trie undergoes lesser number of
burst operations, and the total insertion time decreases with increasing
alphabet size.  Figure~\ref{fig:zip_k}(b) shows that the insertion time for the
compact trie, however, increases.  The insertion time for \sem{} depends on the
length of the key and the size of the container and is, therefore, mostly
independent of the alphabet size.

Figure~\ref{fig:ser_k}(a) shows the searching time for different alphabet sizes.
For a small alphabet ($k = 2$), the false positive rate is practically $1$ and
the container sizes are extremely large.  As a result, the searching time is
large.  When the alphabet size increases, this probability decreases, thereby reducing the
searching time.  However, for large alphabet sizes, the size of the containers
increase.  Consequently, after $k = 10$, the structures show an increase in the
searching time.

The probability that a key which is absent in the database will still be
searched in a container is given by Eq.~\eqref{eq:entire}.  From the equation,
we can see that more the size of the alphabet is, the lesser is the false
positive rate.  Intuitively, with more characters to choose from, there is a
lesser chance that the same triplet will be randomly chosen by a key in the
database.  Eq.~\eqref{eq:entire_bound} indicates that the amount of pruning
should increase exponentially, and this is validated by
Figure~\ref{fig:ser_k}(b).  Thus, the time for unsuccessful searches decreases
when the alphabet size is increased.  The effect is less prominent for the
direct search strategy as it prunes only on the basis of the last triplet in a
key.

\subsection{Effect of query length on prefix and suffix search}

\begin{figure}[t]
	\begin{center}
		\begin{tabular}{cc}
		\hspace*{-3mm}
		\includegraphics[width=\figwidth]{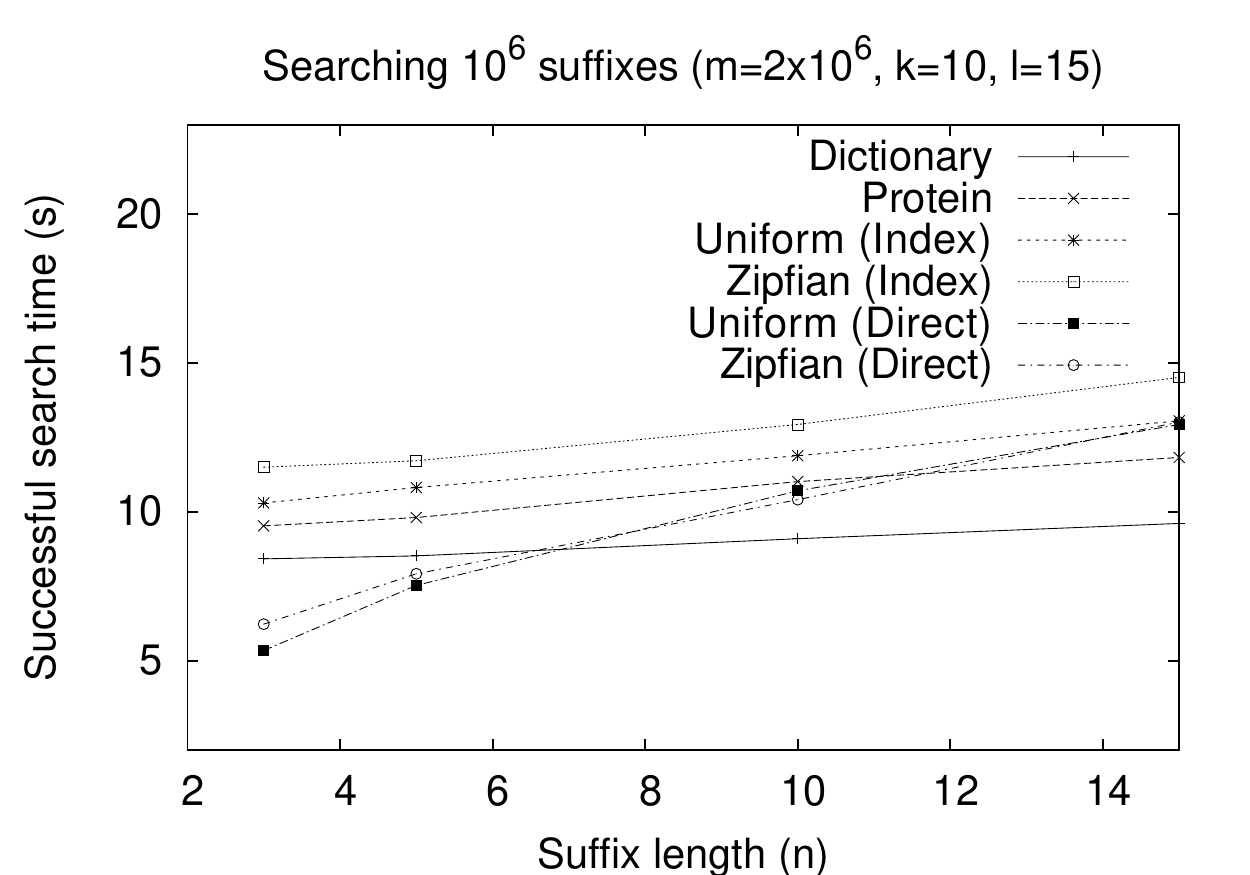} &
		\hspace*{-6mm}
		\includegraphics[width=\figwidth]{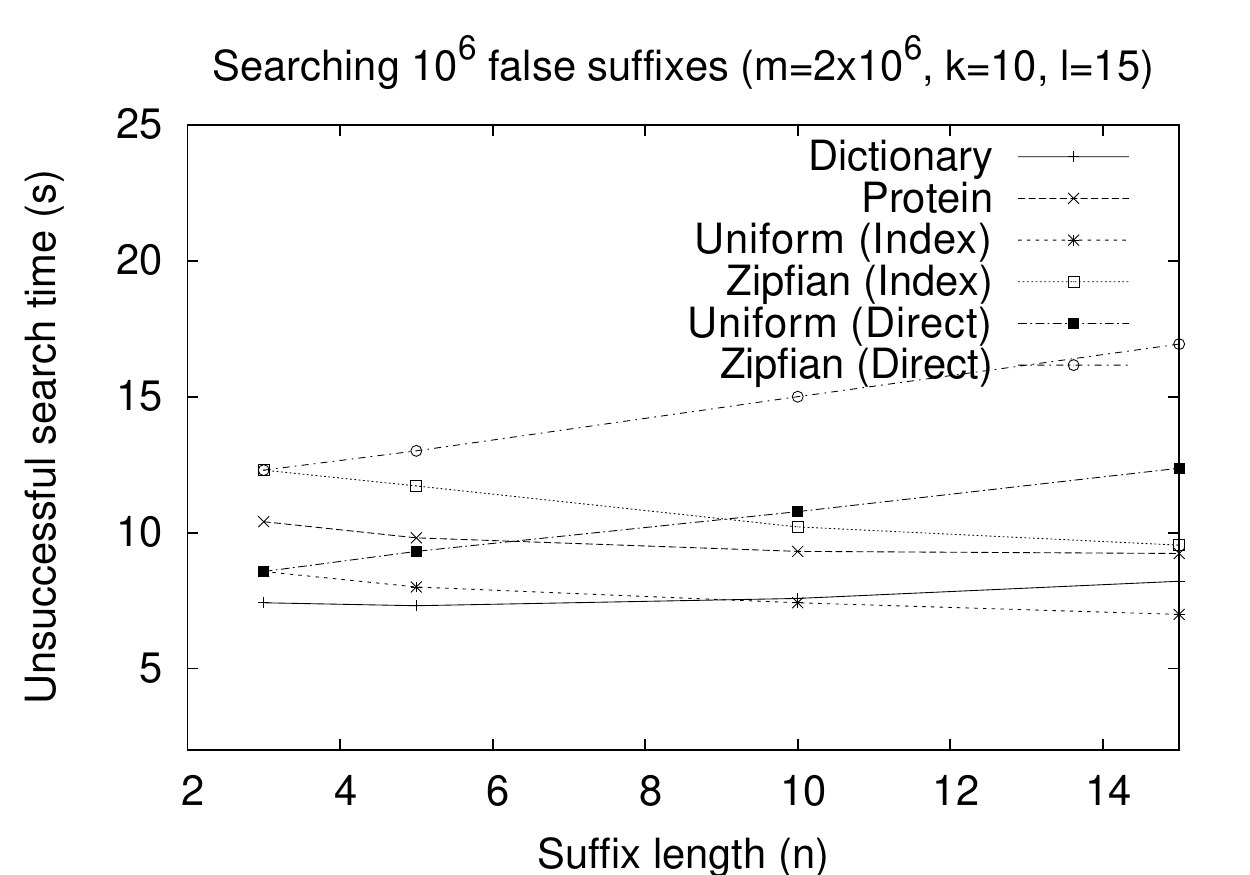} \\
		(a) & (b)
		\end{tabular}
		\caption{Effect of query length on (a) successful and (b) unsuccessful suffix search time.}
		\label{fig:src_suf}
	\end{center}
\end{figure}

\begin{figure}[t]
	\begin{center}
		\begin{tabular}{cc}
		\hspace*{-3mm}
		\includegraphics[width=\figwidth]{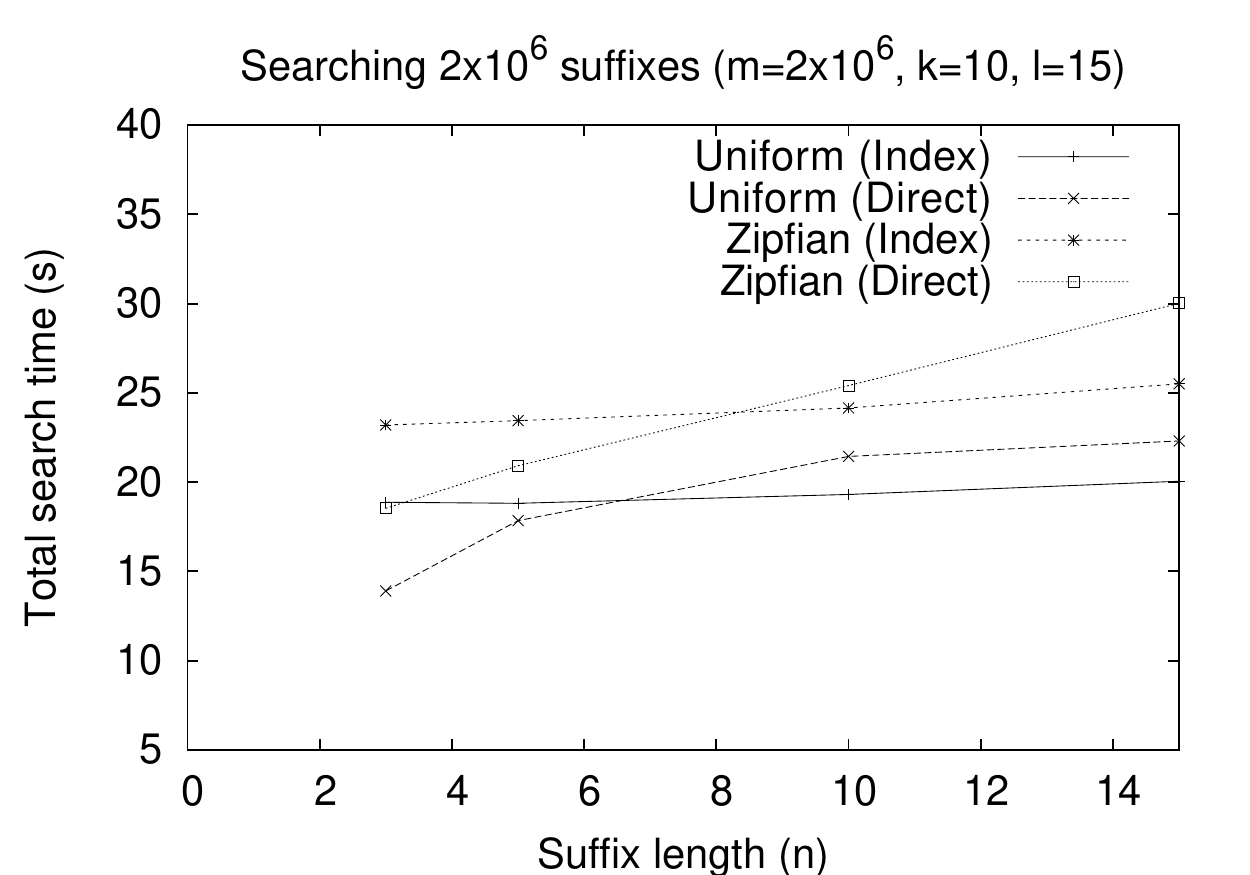} &
		\hspace*{-6mm}
		\includegraphics[width=\figwidth]{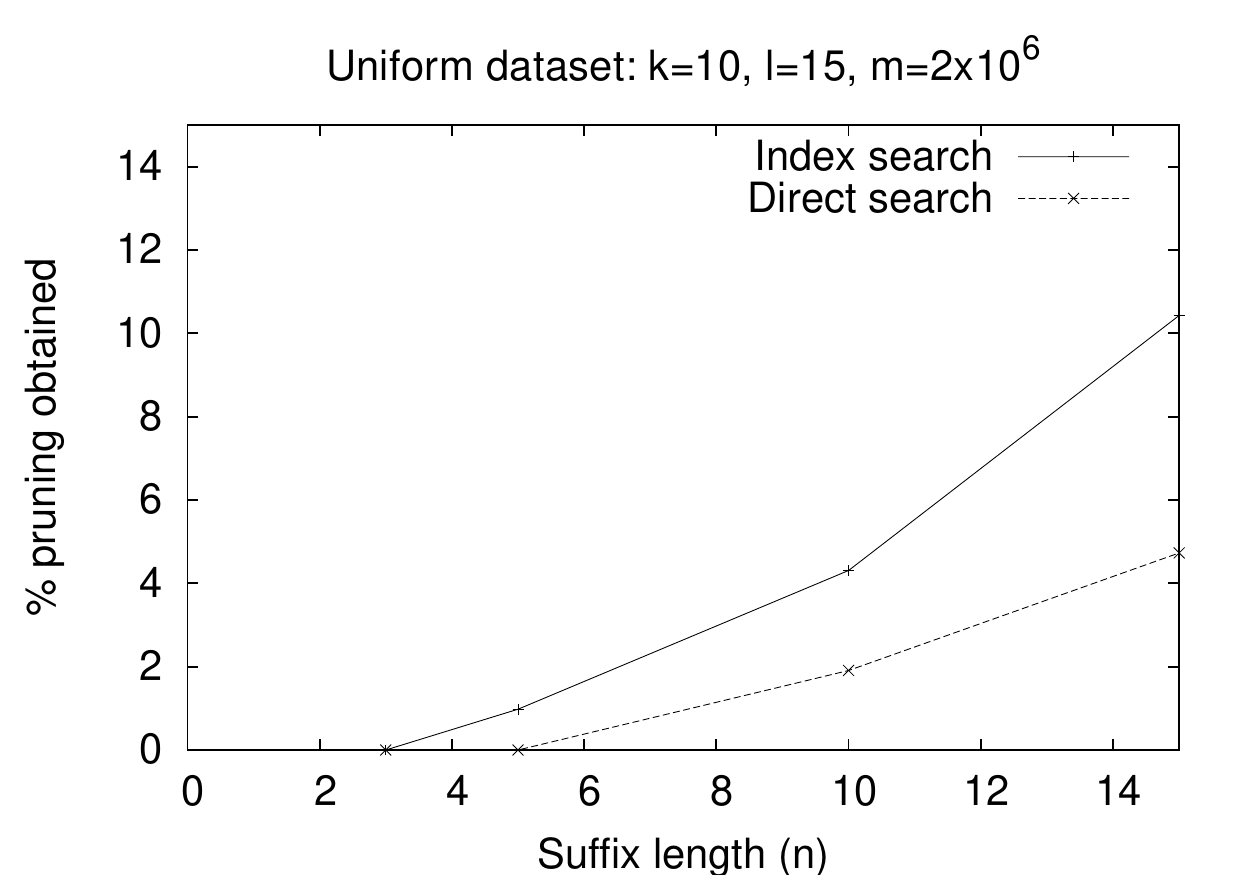} \\
		(a) & (b)
		\end{tabular}
		\caption{Effect of query suffix length on (a) total search time and (b) pruning.}
		\label{fig:prun_suf}
	\end{center}
\end{figure}

The first set of experiments measure the running times for successful,
unsuccessful and total search time for query suffixes of different lengths.
When the presence of the suffix in \sem{} is guaranteed, the direct search
performs better than index search, as it bypasses the overhead of traversing
through the entire length of the query suffix, as indicated by
Figure~\ref{fig:src_suf}(a).  However, for unsuccessful searches, as the length 
of the query suffix increases, the number of triplets increases, producing 
a better pruning ratio for the indexed strategy.  Thus, it performs better as 
shown in Figure~\ref{fig:src_suf}(b).  
Figure~\ref{fig:prun_suf}(a) shows the total search time when both types of
searches are issued.  Overall, the index strategy performs better for larger
query lengths.  

The prefix search experiments showed similar behavior and are, therefore, not
reported.  The effect of the other parameters are roughly equal as that of an
exact key search (refer~\cite{thesis}).

\subsection{Effect of query length on substring search}

\begin{figure}[t]
	\begin{center}
		\begin{tabular}{cc}
		\hspace*{-3mm}
		\includegraphics[width=\figwidth]{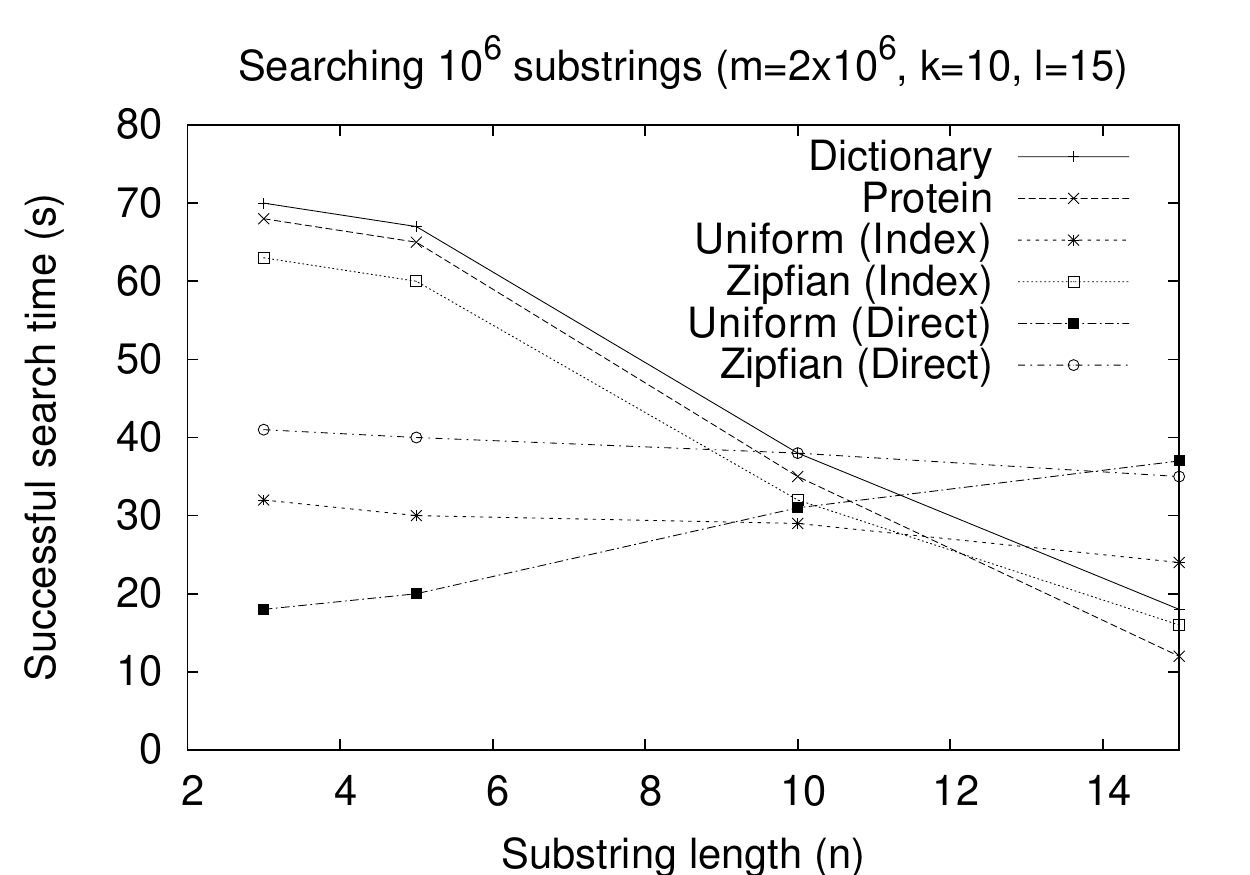} & 
		\hspace*{-6mm}
		\includegraphics[width=\figwidth]{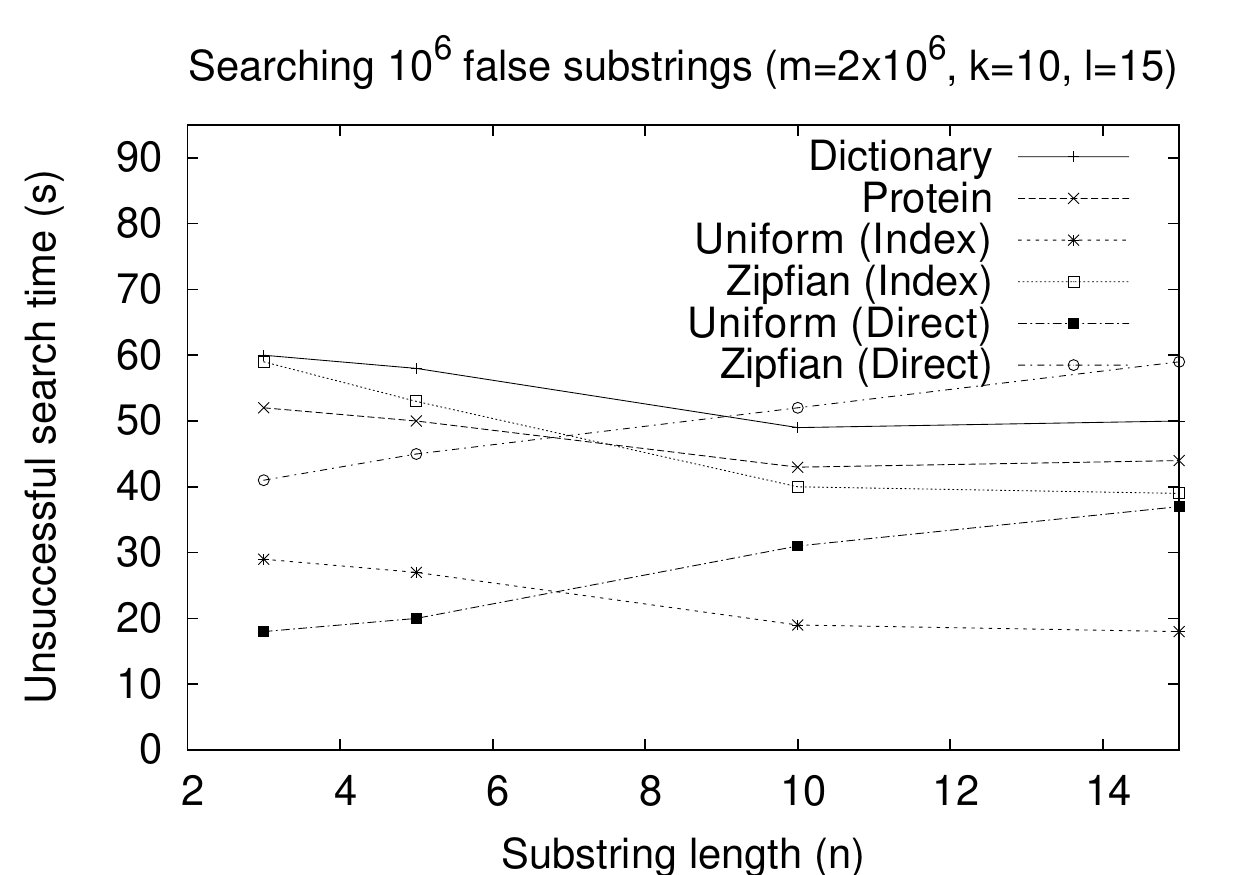} \\
		(a) & (b)
		\end{tabular}
		\caption{Effect of query length on (a) successful and (b) unsuccessful substring search time.}
		\label{fig:src_sbstr}
	\end{center}
\end{figure}

\begin{figure}[t]
	\begin{center}
		\begin{tabular}{cc}
		\hspace*{-3mm}
		\includegraphics[width=\figwidth]{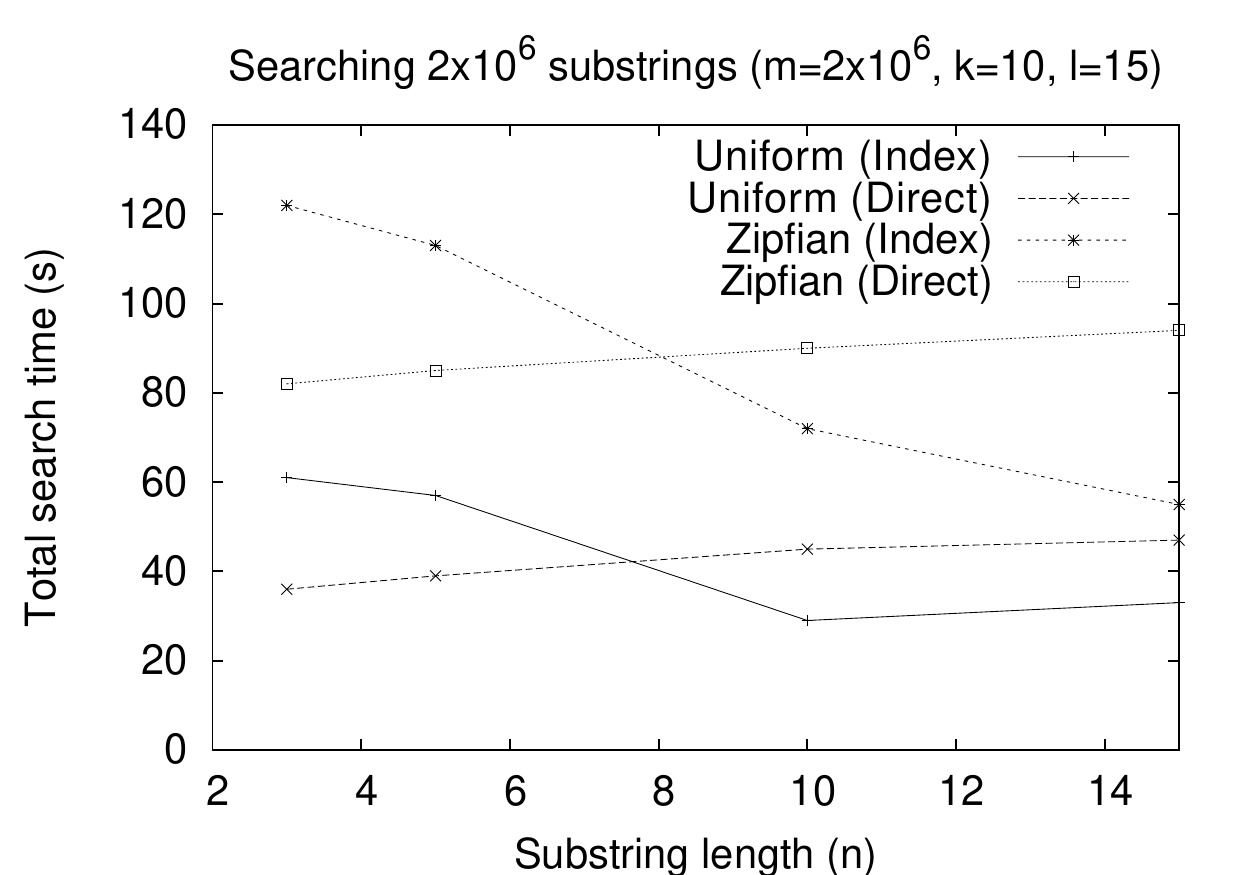} & 
		\hspace*{-6mm}
		\includegraphics[width=\figwidth]{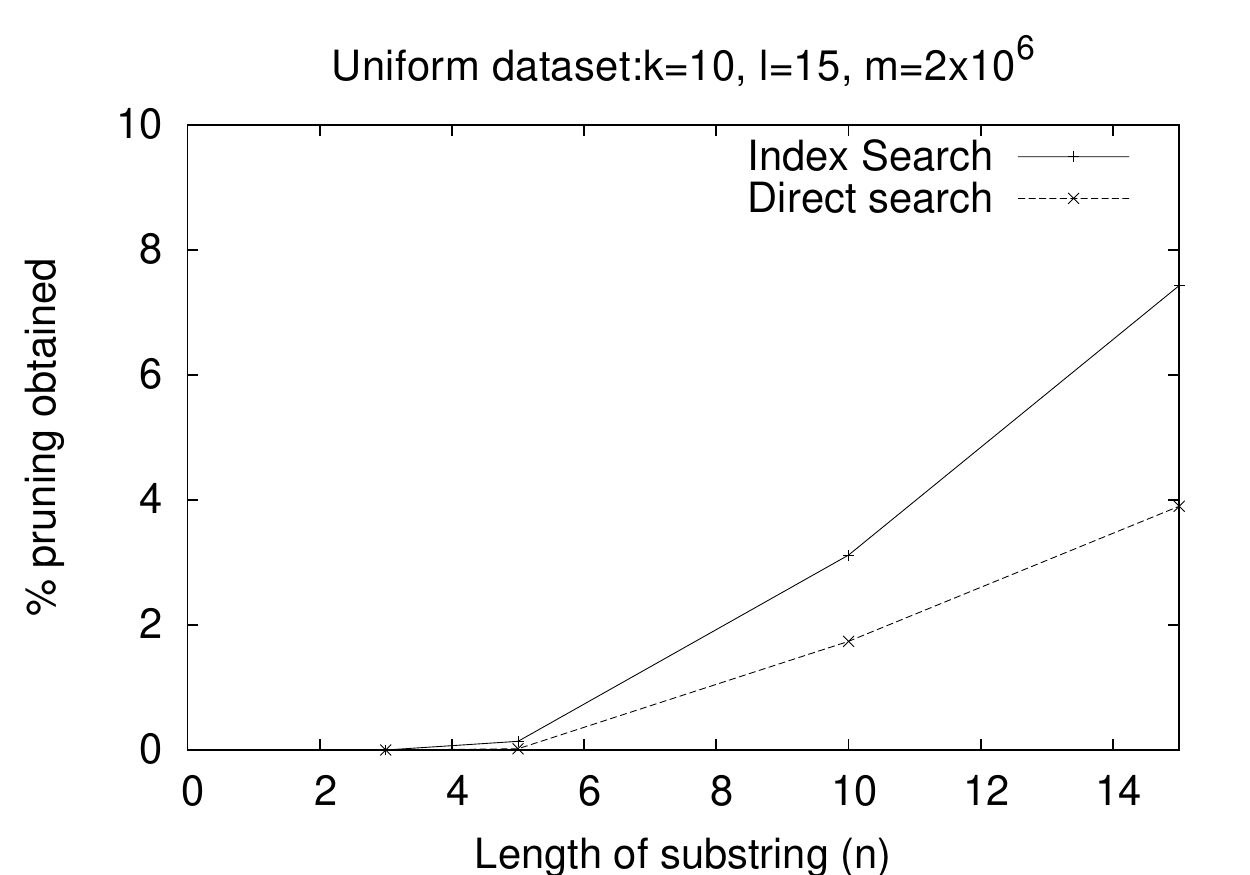} \\
		(a) & (b)
		\end{tabular}
		\caption{Effect of substring length on (a) total search time and (b) pruning.}
		\label{fig:prun_sbstr}
	\end{center}
\end{figure}

The substring search in case of \sem{} involves a collection of prefix search
queries in the additional \sem{} structures.  Hence, the search strategies show
a similar behavior as that of prefix search.  However, as the prefix searches
are done in a number of structures, for a successful substring search, the
direct search will perform much better (Figure~\ref{fig:src_sbstr}(a)) while for
a unsuccessful substring search, the index search will show significant
improvement (Figure~\ref{fig:src_sbstr}(b)) due to the effect of better pruning
of the containers that are searched for larger query substring lengths (as shown in 
Figure~\ref{fig:prun_sbstr}(b)).  The total search time for
both successful and unsuccessful queries is captured in Figure~\ref{fig:prun_sbstr}(a).

\subsection{Summary of experiments}

We can summarize the experimental observations as follows:
\begin{itemize}
	\item Operating on an expanding database, the containers of \sem{} should be
		implemented as a list allowing constant insertion time.  For a
		relatively stable dataset, however, the BST implementation of the
		containers is preferred for efficient retrieval purposes.
	\item For large databases ($10^6$ keys or more), the direct search performs
		better as it does not traverse through the index structure and the
		pruning ratio for both the strategies are almost equal.
	\item When the search query length increases to more than $9$, it is better
		to use the index search strategy as the pruning offered is better.
	\item When the alphabet size is more than $15$, \sem{} is a better choice
		than other structures due to lower memory needs.
\end{itemize}

\section{Conclusions}
\label{sec:concl}

In this paper, we have designed a data structure, \sem{}, that efficiently
manages large sets of strings (or keys) and handles all the different string
queries with low memory requirements.  We
described the indexing technique used by \sem{}, and developed two
variants---list and binary search tree---for the final container of the keys.
We also developed algorithms for different key operations including exact key
searching, insertion, deletion, updating, re-insertion, prefix/suffix searching
and substring searching.  We analyzed how the performance of the different
searching operations and the probability of a search being pruned change with
the number of keys, the length of the key and the alphabet size.  Our
experiments showed that \sem{} is better than the competing structures in terms
of memory size by up to a factor of two, while the insertion and searching times
are either better than or comparable with.

In future, we plan to investigate the effect of modeling the containers as
different data structures such as a hash table, and also how parallelization of the
different procedures improve the running time.

{
\bibliographystyle{abbrv}
\balance
\bibliography{ref}
}

\end{document}